\documentclass[fleqn,usenatbib,useAMS]{mnras}

\usepackage{newtxtext,newtxmath}
\usepackage{braket}
\usepackage{natbib}
\usepackage{booktabs}
\usepackage{multirow}
\usepackage{ulem}
\usepackage[dvipsnames]{xcolor}

\usepackage[T1]{fontenc}

\DeclareRobustCommand{\VAN}[3]{#2}
\let\VANthebibliography\thebibliography
\def\thebibliography{\DeclareRobustCommand{\VAN}[3]{##3}\VANthebibliography}

\usepackage{graphicx}	
\usepackage{amsmath}	

\title[Evolved CVs as AM CVn stars]{Evolved cataclysmic variables as progenitors of AM CVn stars}
\author[Sarkar, Ge \& Tout]{
Arnab Sarkar$^{1}$\thanks{E-mail: as3158@cam.ac.uk},
Hongwei Ge$^{2,1,3}$\thanks{E-mail: gehw@ynao.ac.cn},
and Christopher A. Tout$^{1}$\thanks{E-mail: cat@ast.cam.ac.uk}
\\
\\
$^{1}$Institute of Astronomy, The Observatories, Madingley Road, Cambridge CB3 OHA, UK
\\
$^{2}$Yunnan Observatories, Chinese Academy of Sciences, Kunming 650216, China
\\
$^{3}$University of Chinese Academy of Sciences, Beijing 100049, China
}

\date{Accepted XXX. Received YYY; in original form ZZZ}

\pubyear{2015}

\begin{document}
\label{firstpage}
\pagerange{\pageref{firstpage}--\pageref{lastpage}}
\maketitle

\begin{abstract}
We model cataclysmic variables (CVs) with solar metallicity donors ($X=0.7,\:Z=0.02$) that evolve to form AM CVn stars through the Evolved CV formation channel using various angular momentum loss mechanisms by magnetic braking ($\mathrm{AML_{MB}}$). We find that the time-scale for $\mathrm{AML_{MB}}$ in our double-dynamo (DD) model is shorter than that of previously used empirical formulae. Owing to the shorter time-scales, a larger parameter space of initial conditions evolves to form AM CVn stars with the DD model than with other models. We perform an analysis of the expected number of AM CVn stars formed through the Evolved CV channel and find about $3$ times as many AM CVn stars as reported before. We evolve these systems in detail with the Cambridge stellar evolution code (\textsc{STARS}) and show that evolved CVs populate a region with orbital period $P_\mathrm{orb}\geq5.5\,\mathrm{hr}$. We evolve our donors beyond their orbital period minimum and find that a significant number become extremely H-exhausted systems. This makes them indistinguishable from systems evolved from the He-star and the White Dwarf (WD) channels in terms of the absence of H in their spectra. We also compare the masses, mass-transfer rates of the donor, and the orbital period with observations. We find that the state of the donor and the absence of H in systems such as YZ~LMi and V396~Hya match with our modelled trajectories, while systems such as CR~Boo and HP~Lib match with our modelled tracks if their actual donor mass lies on the lower-end of the observed mass range. 
\end{abstract}


\section{Introduction} 
AM Canum Venaticorum (AM CVn) stars are binaries with extremely short orbital periods, $10\lesssim P_\mathrm{orb}/\,\mathrm{min}\lesssim 65$ \citep{Solheim2010}. \textcolor{black}{They are considered a subclass of cataclysmic variables (CVs), in which a low-mass He-rich star transfers mass to a white dwarf companion. These systems show a general lack of H in their spectra ranging from being partially H-deficient to complete H-deficiency, and are observed at smaller orbital periods than the minimum orbital period of canonical CVs (see \citealt{2003cvs..book.....W} for a detailed review on CVs and their observed properties).} Their evolution is governed primarily by the loss of orbital angular momentum by gravitational radiation and they are also strong sources of gravitational waves \citep{Kupfer2016}.

Three possible formation channels for \textcolor{black}{AM CVn stars} have been \textcolor{black}{proposed}. In the first, known as the White Dwarf (WD) channel, a He WD transfers mass to a more massive \textcolor{black}{carbon-oxygen} (C/O) WD after going through two Common Envelope (CE) phases (see \citealt{2013A&ARv..21...59I} for a review of the CE evolution). Because of the tight orbit of the system, its evolution is primarily governed by angular momentum loss \textcolor{black}{by} gravitational wave radiation ($\mathrm{AML_{GR}}$, see \citealt{Chen2022}, \citealt{Deloye2007} and references therein). However, results of \cite{Shen2015} suggest that all such binary WDs should merge \textcolor{black}{rather than} form \textcolor{black}{AM CVn stars}. The second channel is known as the He-star channel, wherein a non-degenerate He-burning donor star transfers mass to a WD after going through two CE events. \textcolor{black}{Again, the evolution of the system is governed by $\mathrm{AML_{GR}}$ (see \citealt{Yungelson2008} and references therein). However, the results of \cite{2023MNRAS.519.2567S} suggest that other AML mechanisms also contribute to the evolution of systems with $P_\mathrm{orb}\gtrsim20\,\mathrm{min}$.} The final channel is known as the Evolved CV channel \textcolor{black}{in which,} after a \textcolor{black}{single} CE event, a H-rich star commences mass-transfer to a WD accretor \textcolor{black}{while} in the Hertzprung gap (between the end of its main sequence and the beginning of its \textcolor{black}{ascent of the} red giant branch). Such systems \textcolor{black}{can} end up as ultracompact binaries with $P_\mathrm{orb,min}\approx10\,\mathrm{min}$. \textcolor{black}{Their} donors are initially non-degenerate but increasingly become degenerate and H-deficient during \textcolor{black}{their} evolution \citep{Podsiadlowski2003}. For $P_\mathrm{orb}\gtrsim3\,\mathrm{hr}$ the evolution of the system is governed primarily by an AML \textcolor{black}{that probably arises from} magnetic braking in the donor star ($\mathrm{AML_{MB}}$), whereas for shorter periods $\mathrm{AML_{GR}}$ drives the evolution  (see \citealt{Solheim2010} for a thorough discussion on all possible formation channels).

The Evolved CV formation channel is usually given less importance primarily because many observed \textcolor{black}{AM CVn stars} do not have traces of hydrogen in their spectra and it is claimed that the Evolved CV channel would \textcolor{black}{always} leave traces of hydrogen in the system (see \citealt{Nelemans2010} and the references therein). In addition, the relative importance of this formation channel has been questioned by the work of \cite{2004MNRAS.349..181N} who find that extensive fine-tuning of initial conditions and long time-scales are required to remove all hydrogen from the system. \textcolor{black}{However it is important to mention that some known \textcolor{black}{AM CVn stars} have been suggested to have evolved through this formation channel (see \citealt{Solheim2010} and the references therein)}. One of the main uncertainties with the current conclusions about the relative importance of the Evolved CV channel is its strong dependence on the assumed mechanism for $\mathrm{AML_{MB}}$. Previous studies have heavily relied on the empirical magnetic braking formula of \cite{1983ApJ...275..713R} and therefore their conclusions strongly depend on it.  \citet[hereinafter ST]{2022MNRAS.513.4169S}, following \cite{Zangrilli1997}, have come up with a physically motivated formalism for $\mathrm{AML_{MB}}$ in CVs, known as the double dynamo (DD) model, and its \textcolor{black}{partial} cessation when the donor becomes fully convective, to explain the occurrence of the period gap in the observed CV distribution \citep{2003cvs..book.....W}. They report good agreement \textcolor{black}{between} their model of $\mathrm{AML_{MB}}$ for zero-age CVs \textcolor{black}{and} that of \cite{Knigge2011}, who use a modified version of the \textcolor{black}{formula of} \citet[see fig.~11 of ST]{1983ApJ...275..713R}. Here we revisit in detail the Evolved CV channel of AM CVn formation using our DD model. \textcolor{black}{All the results in this work are for solar metallicity.}

In Section \ref{sec:taus}, we compare the different time-scales for $\mathrm{AML_{MB}}$ and conclude that \textcolor{black}{they} are short enough for evolved CVs to end up as \textcolor{black}{AM CVn stars} within $12\:\mathrm{Gyr}$. In Section \ref{sec:probanaly} we show that a larger parameter space of initial conditions \textcolor{black}{leads to} \textcolor{black}{AM CVn stars} with the DD model than with other empirical models. We make detailed models of systems from their initial configuration to beyond their orbital period minimum and compare them with observed systems in Section \ref{sec:model}. We discuss the shortcomings of our analysis and scope for further research in Section \ref{sec:disc}. We summarize and conclude our work in Section \ref{sec:Conclusion}.

\section{The dependence of the secular evolution of CV systems on the AML mechanism}
\label{sec:taus}
\textcolor{black}{In this section we illustrate the dependence of our modelled trajectories on the assumed model of AML. We show that qualitatively different binary systems arise from the same initial conditions when we change the AML mechanism.} We take a system with a donor of mass $M_2=1M_\odot$, a WD accretor of mass $M_1=1M_\odot$ and an initial orbital period $P_\mathrm{orb}=2.3\:\mathrm{d}$ and evolve it \textcolor{black}{with} three different AML mechanisms \textcolor{black}{using the Cambridge stellar evolution code (\textsc{STARS})}. The first is that of \cite{2022MNRAS.513.4169S}. We use their formula for AML by magnetic braking, $\mathrm{AML_{MB}}$, their equation (25), with the three free parameters $(\alpha, \beta,\gamma)=(4.6, 0.08, 3.2)$\footnote{See their section 3 for the expressions and a detailed explanation of all the relevant terms.}. These free parameters were chosen to explain the period gap and the extra AML below the period gap in the CV distribution \citep{2003cvs..book.....W, Gnsicke2009}. We add AML by gravitational radiation, ($\mathrm{AML_{GR}}$, see \citealt{1981ApJ...248L..27P} and the references therein), given by ST's equation (26). We call this model the DD model. The other $\mathrm{AML_{MB}}$ mechanism we use is that of \citet{1983ApJ...275..713R} given by equation (28) of ST, with $\delta=4$, along with the same expression for $\mathrm{AML_{GR}}$ as in the DD model. We call this the RVJ model. The last model we use is a modification to the RVJ model given by \citet{Knigge2011} in order to explain the period gap and the extra AML below the period gap. Equation (28) of ST is multiplied by a factor of 0.66 with $\delta=3$ and equation (26) is multiplied by a factor of 2.47. We call this the KBP model.

\begin{figure}
\includegraphics[width=0.5\textwidth]{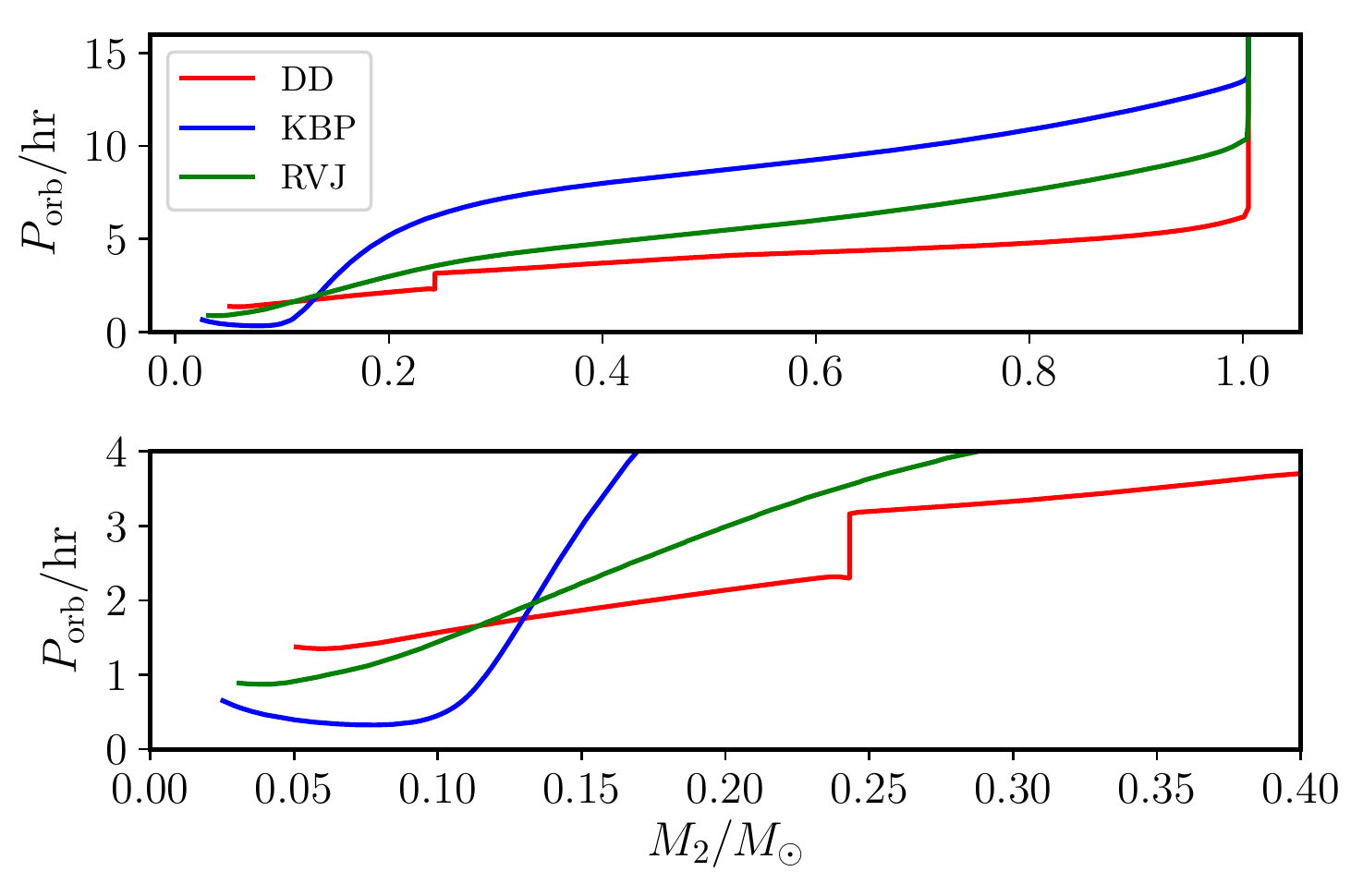}
\caption{The evolution in the ($M_2,\,P_\mathrm{orb}$) plane of a system with \textcolor{black}{initially} $M_2=1M_\odot$,  $M_1=1M_\odot$ and $P_\mathrm{orb}=2.3\:\mathrm{d}$ for three different \textcolor{black}{rates} for $\mathrm{AML_{MB}}$. The DD model produces a canonical CV with a period gap, whereas RVJ and KBP models produce \textcolor{black}{AM CVn stars}. \textcolor{black}{The systems become CV, (canonical or evolved) after the commencement of RLOF and evolve from right to left}.}
\label{fig:pm_new}
\end{figure}

\begin{figure}
\includegraphics[width=0.5\textwidth]{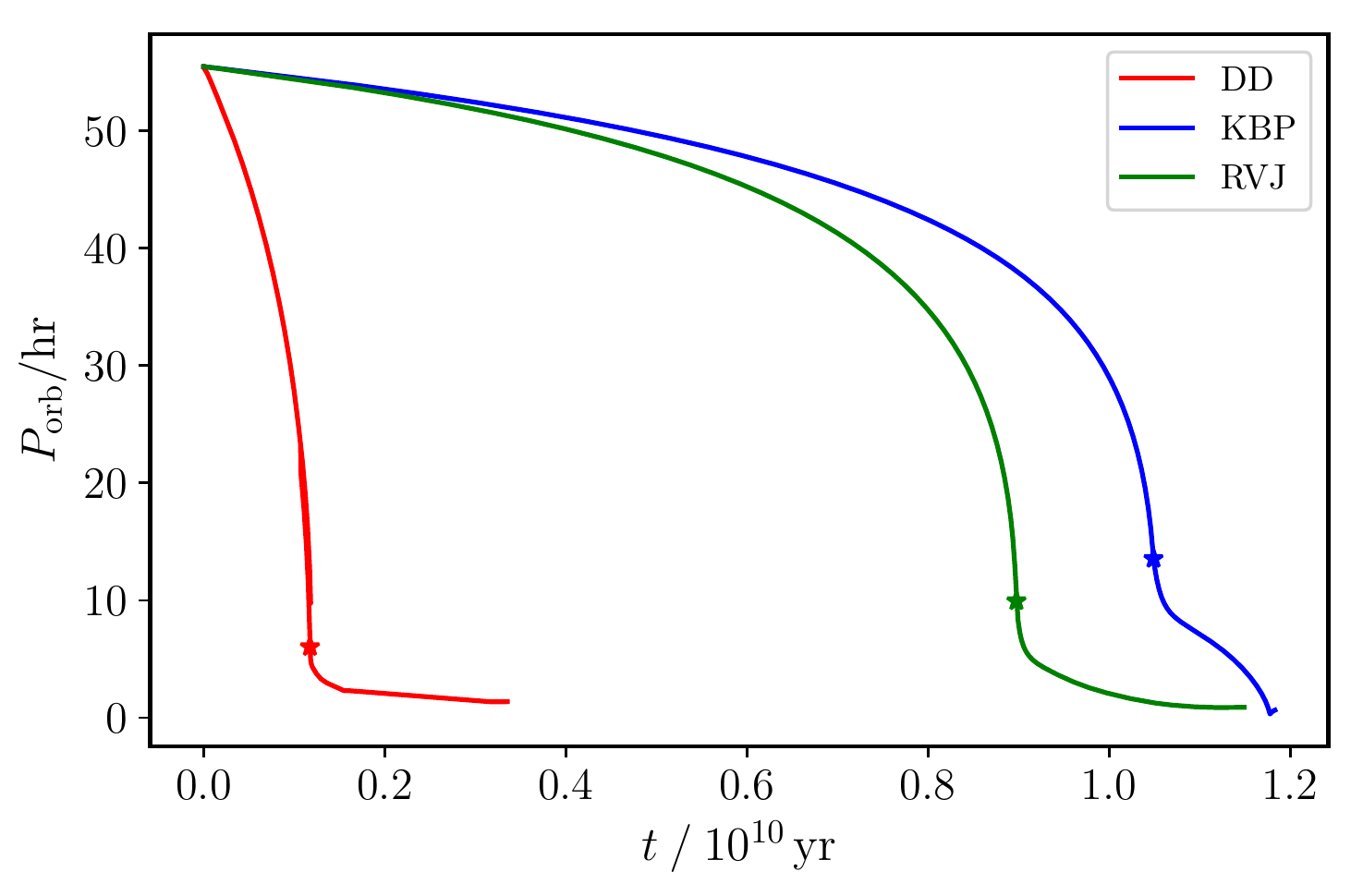}
\caption{The evolution of $P_\mathrm{orb}$ with time for the system in Fig.~\ref{fig:pm_new}. \textcolor{black}{The stars indicate the orbital period and the time of commencement of RLOF.} The system evolves much faster with the DD $\mathrm{AML_{MB}}$ model than with the other two models.}
\label{fig:pt_new}
\end{figure}

Our results are shown in Figs~\ref{fig:pm_new} and \ref{fig:pt_new}. Fig.~\ref{fig:pm_new} shows the evolution of the donor following Roche lobe overflow (RLOF), whence mass loss to the accretor drives the evolution from right to left. It can be seen that the DD model leads to the system evolving as a canonical CV, with a period gap in the region $2\lesssim P_\mathrm{orb}/\mathrm{hr}\lesssim3$ and a period minimum of $P_\mathrm{orb, min}\approx1.5\,\mathrm{hr}$. However the RVJ model leads to the disappearance of the period gap and a smaller $P_\mathrm{orb, min}$, so forming an AM CVn system and demonstrating that the donor is more H-exhausted than that in a canonical CV, \textcolor{black}{in accordance with the results of \cite{1989A&A...208...52P} which state that CVs with evolved donors either do not produce a period gap or have a smaller period gap.} The donor in the KBP model is even more H-exhausted \textcolor{black}{resulting in} an even smaller $P_\mathrm{orb, min}$. \textcolor{black}{This is because compared to the DD AML, the evolution governed by the empirical AMLs (RVJ and KBP) leads to a slower loss of angular momentum, and consequently RLOF commences when the donor is more H-exhausted. In general, $\mathrm{AML_{MB}}$ time-scale ($\tau_\mathrm{MB} = J/\Dot{J}_\mathrm{MB}$) and consequently the time-scale of mass loss ($\tau_\mathrm{ML} = M_2/\Dot{M}_2$) is shortest for the DD model, followed by the RVJ model and the KBP model, leading to a quicker shrinkage of the Roche lobe and consequently earlier mass transfer by RLOF. Overall, the evolution of the system proceeds much faster with the DD model than the other two, as shown in Fig.~\ref{fig:pt_new}. The DD model leads to RLOF during the main-sequence (MS) phase of the donor, whereas the RVJ and the KBP model lead to RLOF beyond the end of MS.} This \textcolor{black}{illustrates} how sensitive the trajectory of CVs is to the assumed $\mathrm{AML_{MB}}$ mechanism. 

\section{AM CVn stars through the Evolved CV channel}
\label{sec:probanaly}
The Evolved CV channel requires only a single CE phase, after which the orbital separation is short enough for the secondary to overfill its Roche lobe as a MS or subgiant star, rather than a more evolved giant.
If the donor in such systems begins RLOF at the end of the MS, the system ends up as an ultracompact CV with $P_\mathrm{orb,min}$ ranging from \textcolor{black}{less than about} $60\,\mathrm{min}$ to as short as $10\,\mathrm{min}$ \citep{Solheim2010}. The system evolves to shorter $P_\mathrm{orb,min}$ the more H-exhausted the donor is at the commencement of RLOF (see also Section \ref{sec:model}). However, if RLOF \textcolor{black}{begins on} the Red-Giant branch (RGB), the system may instead evolve to form a wide binary (see fig.~1 of \citealt{2016ApJ...833...83K} and the references therein). This critical initial separation (or initial $P_\mathrm{orb}$) that separates the ultracompact converging CVs from the wide diverging CVs is known as the bifurcation limit \citep{1988A&A...191...57P}. \textcolor{black}{In order to quantitatively analyse the likelihood of systems evolving as AM CVn stars, we assume for the rest of the paper that in order to form \textcolor{black}{AM CVn stars} RLOF should \textcolor{black}{begin} when the donor is between its \textcolor{black}{terminal main sequence} (TMS) phase and RGB phase.} With this assumption, we describe a simple method to obtain progenitors of \textcolor{black}{AM CVn stars} by this channel.

Before the commencement of RLOF ($\Dot{M}_1=\Dot{M}_2=0$) we can write




\begin{center}
\begin{equation}
\label{eq:j3}
\frac{\Dot{R}_\mathrm{L}}{R_\mathrm{L}} = 2\frac{\Dot{J}}{J},
\end{equation}
\end{center}
where $J$ is the orbital angular momentum of the system and $R_\mathrm{L}$ is the Roche lobe radius of the secondary. This is because $J\propto a^{1/2}$ and $R_L\propto a$. Thus we can evolve the system until RLOF (where $R_\mathrm{L}=R_2$) easily, if provided with the initial $M_2$, $M_1$ and $a_\mathrm{initial}$, using a simple \textcolor{black}{Euler integration} for $a$ and consequently $R_\mathrm{L}$, given by

\begin{figure}
\includegraphics[width=0.5\textwidth]{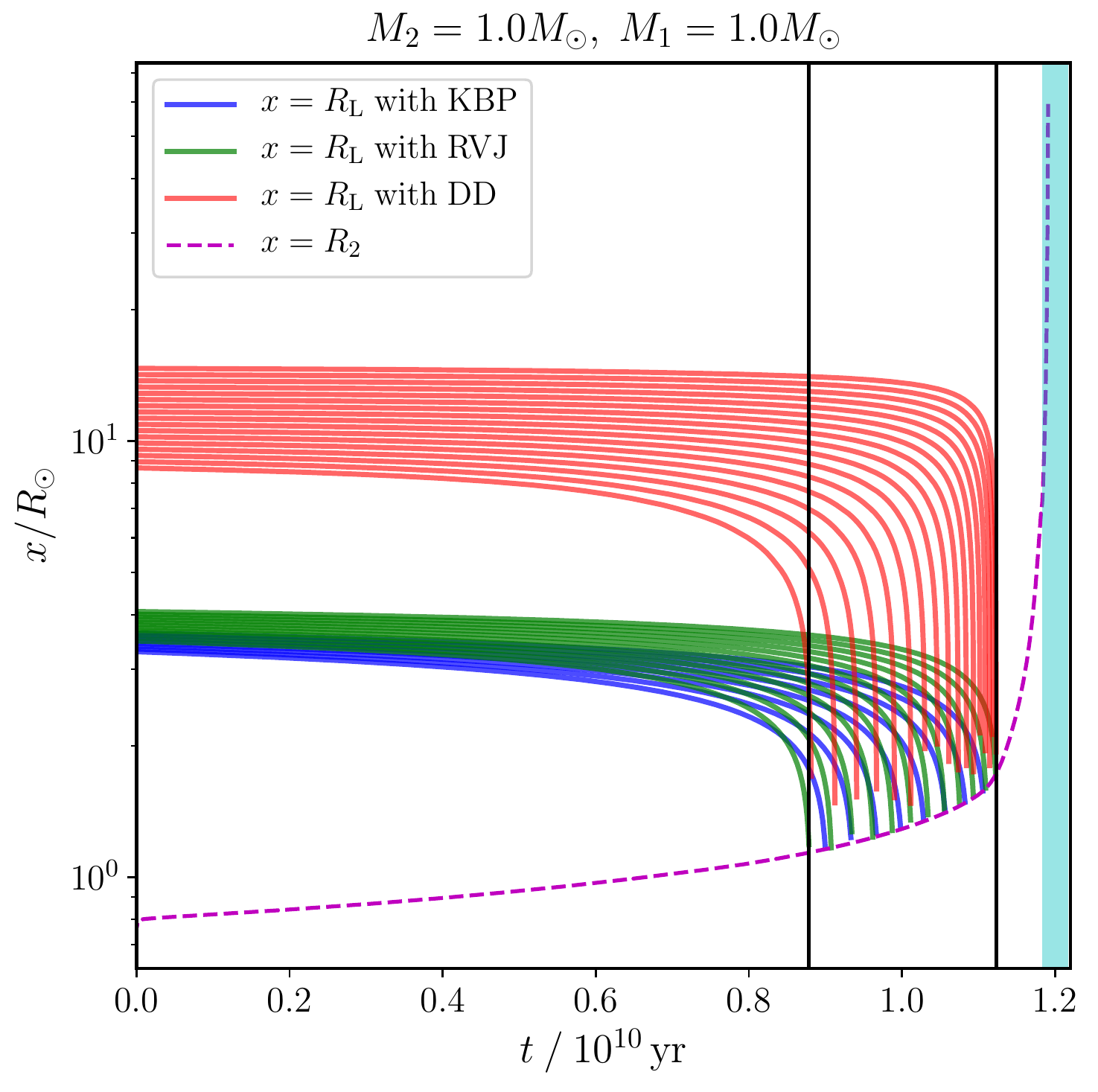}
\caption{A plot showing the evolution of various parameters with time \textcolor{black}{until the commencement of RLOF}. \textcolor{black}{ The red trajectories are the various tuples ($M_2=1M_\odot,\,M_1=1M_\odot,\,R_\mathrm{L,\,initial}$), equidistant in log$\,a$, of the DD model with $t_\mathrm{TMS}\leq t_\mathrm{RLOF}\leq t_\mathrm{BGB}$. Similarly, the green and the blue trajectories are the tuples that satisfy this criterion for the RVJ and KBP models. The magenta curve is the evolution of the donor radius with time. The vertical black lines denote $t_\mathrm{TMS}$ and $t_\mathrm{BGB}$, and the cyan region is the age of the Galaxy, which we take to be 12 Gyr.}}
\label{fig:rl_ev_egg}
\end{figure}

\begin{center}
\begin{equation}
\label{eq:a}
a_{t+\Delta t} \approx a_t + \Dot{a}_t(a_t)\Delta t,
\end{equation}
\end{center}
where $a_\mathrm{initial} = a_{t=0}$ and $\Dot{a}_t$ is itself a function of $a_t$. Evolving the RVJ and the KBP models is simple. This only requires $R_2$ as a function of time (see equation 28 of ST). \textcolor{black}{However to evolve the DD model we need as a function of time $R_2$, the mass of the convective envelope $M_\mathrm{env}$, the radius of the convective envelope $R_\mathrm{env}$, the donor luminosity $L_2$, the radius of the radiative core $R_\mathrm{core}$, the effective thickness of the boundary layer between the core and the envelope $H_\mathrm{B}$, and the density of this boundary layer $\rho_\mathrm{B}$.} We refer the reader to section 3 of ST for a detailed explanation of the relevance of each of these terms (see also \citealt{Zangrilli1997} for a detailed explanation of the DD model). The evolution of the parameters until RLOF in all three models depends only on the evolution of an isolated star, so we evolve a single star with $X=0.7$, $Y=0.28$ and $Z=0.02$ with the \textsc{STARS} code till it evolves past the base of the RGB (BGB) and obtain all the required parameters for each timestep $\Delta t$. Now we use equation (\ref{eq:a}) to evolve $a$ and $R_\mathrm{L}$ with time. For each timestep we check whether $R_\mathrm{L}\leq R_2$ indicating \textcolor{black}{the onset} of RLOF $t_\mathrm{RLOF}$. If so we check if $t_\mathrm{TMS}\leq t_\mathrm{RLOF}\leq t_\mathrm{BGB}$. \textcolor{black}{If a system satisfies this criterion, the tuple ($M_2,\,M_1,\,a_\mathrm{initial}$) describes the properties of a potential progenitor of \textcolor{black}{AM CVn stars}.} Otherwise we discard the system. We also ensure that RLOF \textcolor{black}{begins} before 12 Gyr, which we take to be the age of the Galaxy
\footnote{For RVJ and KBP, the full evolution may take longer than 12 Gyr even though RLOF \textcolor{black}{begins} much earlier. This is not a problem for the DD model because of its shorter evolutionary time-scale.}. In order to obtain estimates of $t_\mathrm{TMS}$ and $t_\mathrm{BGB}$ we use the turning points of \citet[table~1, models C and E respectively]{1998MNRAS.298..525P}~\footnote{\textcolor{black}{However, we note that these transition points are not well-defined in low-mass stars and so our results are sensitive to changes in these values.}}. Our result is shown in Fig.~\ref{fig:rl_ev_egg} for a $M_1=1M_\odot$ and $M_2=1M_\odot$ system. It can be seen that systems in a larger \textcolor{black}{range} of log$\,R_\mathrm{L}$ and thus log$\,a$ can potentially evolve to \textcolor{black}{AM CVn stars} with the DD model than with either the RVJ or KBP model.

\begin{figure}
\includegraphics[width=0.5\textwidth]{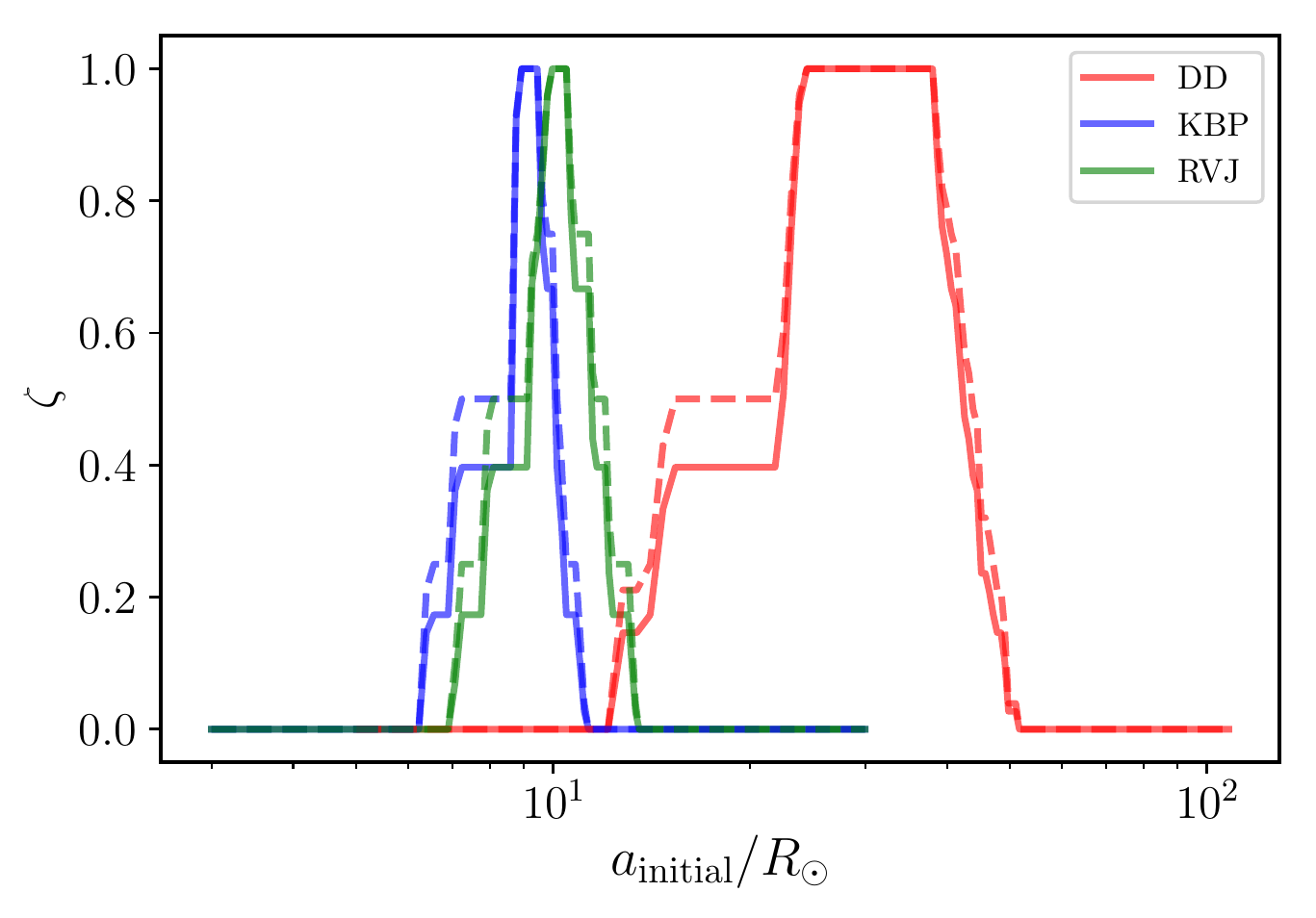}
\caption{\textcolor{black}{The weighted fraction $\zeta$, as a function of $\mathrm{log}\,a_\mathrm{initial}$, of our 28 simulated systems (see text) that form \textcolor{black}{AM CVn stars} for different $\mathrm{AML_{MB}}$ mechanisms.} \textcolor{black}{The solid lines are for $f_2(m)$ modelled with the Salpeter IMF, while the dashed lines are for $f_2(m)$ modelled with a flat distribution.}}
\label{fig:prob}
\end{figure}

We now apply the same method to a set of donor stars with $M_2/M_\odot \in \{1.0,1.1,1.2,1.3\}$ and with WD masses $M_1/M_\odot \in \{0.6,0.7,0.8,0.9,1.0,1.1,1.2\}$, leading to 28 different combinations of donors and accretors. None of these systems have $q=M_2/M_1$ greater than the critical ratio for dynamical mass transfer \citep{Ge2015, Ge2020}. The choice of donor masses was motivated by the analysis of \cite{Podsiadlowski2003}. We do not consider donors with masses less than $1\,M_\odot$ because for them $t_\mathrm{TMS}>12\:\mathrm{Gyr}$. We also do not consider donors with $M_2>1.3\,M_\odot$ because they develop a radiative envelope, making the DD model of ST inefficient, because its formulation requires a convective envelope for $\mathrm{AML_{MB}}$. Another reason for limiting our analysis to $M_2\leq1.3\,M_\odot$ is that at the end of this section we attempt a rough scaling of the birthrate \textcolor{black}{similar to} \cite{Podsiadlowski2003} who use donors with $M_2\leq1.4\,M_\odot$. The choice of accretors is motivated by robust estimates of the observed mass of WDs in CVs \citep{Wijnen2015}. We find the region in $a_\mathrm{initial}$ which gives rise to \textcolor{black}{AM CVn stars} for each of these 28 models \textcolor{black}{with} the method described (see Appendix~\ref{ap:tab}). \textcolor{black}{Now we obtain a single scaled distribution for the different AML models, for which we require a mass distribution function for the donor secondaries, as well as a distribution function for orbital separation of the AM CVn progenitors. The donors of AM CVn stars are the less-massive component of the primordial MS-MS binary, the more massive component of which underwent a CE event that formed the WD accretor. The Evolved CV channel differs from the other two channels such that system parameters of the WD plus secondary star system avoided a second CE phase and began mass-transfer in a dynamically stable way. The two events that lead up to the formation of the mass-transferring H-deficient secondary plus the mass-accreting WD, namely undergoing the first CE phase and avoiding the second CE phase significantly alter the mass distribution function of the donor secondary and the orbital separation distribution of the system, whence a simple IMF treatment such as using the Salpeter IMF \citep{Salpeter1955} or a flat distribution in $q$ is incorrect. Similarly, it is incorrect to assume a flat distribution in $\mathrm{log}\,a_\mathrm{initial}$ because of a reported correlation between the orbital separation and $q$ \citep{2017ApJS..230...15M}. Modelling these effects is quite difficult and beyond the scope of this work, so we work with simple distributions for the two parameters. We set a uniform distribution in $\mathrm{log}\,a_\mathrm{initial}$, and analyse two secondary mass distribution functions $f_2(m)$, the first being the Salpeter IMF, and the other being a uniform distribution. The two $f_2(m)$ show two extreme effects of choosing the mass distribution function of the secondaries, wherein the former implies no dependence of the secondary mass distribution on the two CE phases, while the latter shows the likely effect of the two CE events with an extremely shallow-sloped mass distribution function compared to the Salpeter IMF. We multiply our secondaries with $f_2(m)$, and} for the accretors we weigh them according to the WD distribution in CVs \citep[see the black histogram in fig.~6 of ][]{Wijnen2015}. Finally, we scale our distribution to obtain $\zeta$, 
\textcolor{black}{which denotes the fraction of our 28 systems that form \textcolor{black}{AM CVn stars}. If $\zeta=1$ for some $\mathrm{log}\,a_\mathrm{initial}$, all of the 28 systems can form \textcolor{black}{AM CVn stars} if the evolution begins with this $\mathrm{log}\,a_\mathrm{initial}$ (or $\mathrm{log}\,P_\mathrm{initial}$). Similarly, if $\zeta=0$ for this region none of the 28 systems can evolve to form \textcolor{black}{AM CVn stars}.} This is shown in Fig.~\ref{fig:prob}. It can be seen that a larger separation space can become AM CVn progenitors with the DD model than with the other two models when we assume a flat distribution in $\mathrm{log}\,a_\mathrm{initial}$ \footnote{\textcolor{black}{This assumption is not strictly valid for such post-CE binaries, as discussed in Section~\ref{sec:disc}.}}. In particular, the area under the curve for the DD model in the $(\mathrm{log}\,a_\mathrm{initial},\zeta)$ plane is about 3 times that for the RVJ model \textcolor{black}{for both choices of $f_2(m)$}. However we note very importantly that, owing to the long time-scales of $\mathrm{AML_{MB}}$ (Fig.~\ref{fig:pt_new}), not all systems which commence RLOF between $t_\mathrm{TMS}$ and $t_\mathrm{BGB}$ necessarily form \textcolor{black}{AM CVn stars} within $12\,\mathrm{Gyr}$ with the RVJ and the KBP models. This further reduces the parameter space of AM CVn progenitors in these models. We obtain a rough estimate of the revised number of expected \textcolor{black}{AM CVn stars} formed through the Evolved CV channel by following the calculation made by \citet[see their section~6]{2004MNRAS.349..181N}. The average time spent by a system with $M_2=1M_\odot$ and $M_1=1M_\odot$ such that its $P_\mathrm{orb}\leq 25\,\mathrm{min}$ is about $ 2.88\times 10^7\:\mathrm{yr}$. Multiplying this by the Galactic birthrate of $6.4\times10^{-5}\:\mathrm{yr^{-1}}$ from \cite{Podsiadlowski2003} and multiplying this by $3$ to account for the increased parameter space  of AM CVn formation with the DD model, we get an expected number of ultracompact CVs with $P_\mathrm{orb}\leq 25\,\mathrm{min}$ to be about $5700$ \textcolor{black}{with the DD model}, as opposed to 2400 obtained by \cite{2004MNRAS.349..181N}. However, this number is certainly an underestimate because many observed \textcolor{black}{AM CVn stars} have $P_\mathrm{orb}\geq 25\,\mathrm{min}$ (see Figs~\ref{fig:pm-solvan} and \ref{fig:mdotP-solram}). So we repeat the calculation for systems with $P_\mathrm{orb}\leq 45\,\mathrm{min}$ and find the expected number of such systems to be greater than $ 93100$\footnote{This number is still an underestimate because our code fails to fully track the most H-exhausted systems beyond the orbital period minimum when the orbit expands as a result of mass loss.}.

\section{Detailed models of AM CVn systems}
\label{sec:model}
\begin{figure}
\includegraphics[width=0.5\textwidth]{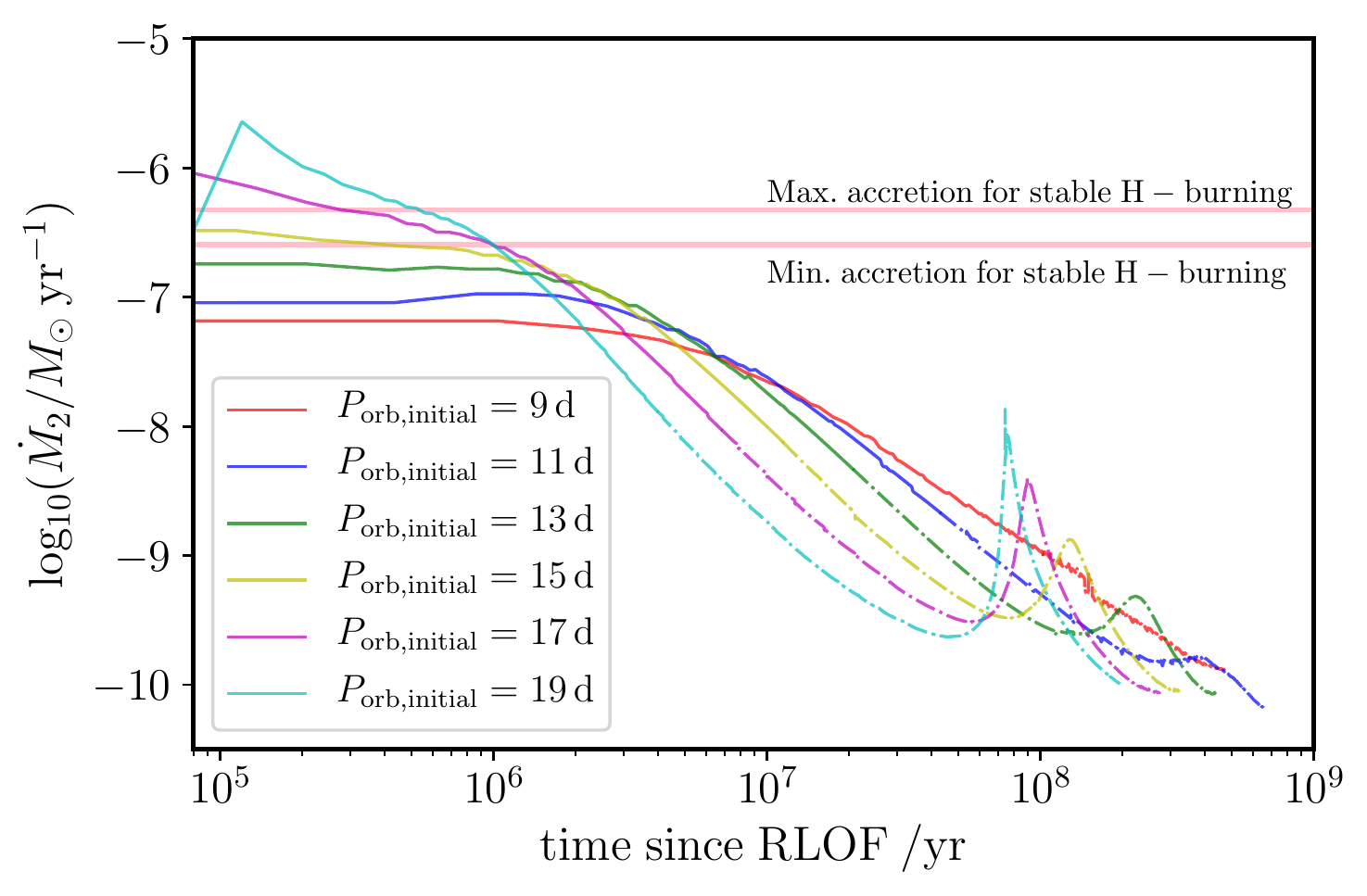}
\caption{The evolution of $\Dot{M}_2$ with time t since RLOF began in the DD model for six systems with $M_2=1M_\odot$, $M_1=1M_\odot$ and various $P_\mathrm{orb,initial}$ such that $P_\mathrm{orb,initial}=9\,\mathrm{d}$ and $P_\mathrm{orb,initial}=19\,\mathrm{d}$ commence RLOF at around $t_\mathrm{TMS}$ and $t_\mathrm{BGB}$ respectively. Solid lines represent H-dominated matter and the dash-dot lines represent He-dominated matter. The horizontal lines are the maximum and and minimum accretion rates for stable hydrogen burning on the surface of the WD accretor  \protect\citep{Wang2018}.}
\label{fig:mdot-t}
\end{figure}
In this section, we make detailed models of the systems that form \textcolor{black}{AM CVn stars} using \textsc{STARS}. We model donor stars from $(M_{2,t=0},M_{1,t=0},P_\mathrm{orb,initial})$ till the code \textcolor{black}{fails} beyond the system's orbital period minimum\footnote{This is because, in the current version of \textsc{STARS}, mass cannot be removed from a degenerate region. We shall fix this in the near future.}. Each donor has $X=0.7$, $Y=0.28$ and $Z=0.02$. We assume that the system undergoes non-conservative mass transfer from the donor to the accretor such that all the mass accreted onto the WD is expelled in nova eruptions and carries away the specific angular momentum of the WD. We test the validity of this assumption in Fig.~\ref{fig:mdot-t} specifically for the DD model\footnote{The RVJ and the KBP models always have $\Dot{M}_2\lesssim10^{-8}M_\odot\,\mathrm{yr^{-1}}$.} because of its high mass-transfer rate at the beginning of RLOF. We compare the mass-transfer rates of our systems with the critical mass-accretion rate of WDs for \textcolor{black}{stable H-burning} from \citet[see their equations 1 and 2]{Wang2018}. It can be seen that the systems spend most of their evolution with $\Dot{M}_2$ below the minimum accretion rate for stable H-burning such that multicycle H-shell flashes, like nova eruptions, can occur on the WD surface by unstable nuclear burning. However, for the two most H-exhausted systems, with $P_\mathrm{orb,initial}=17\,\mathrm{d}$ and $P_\mathrm{orb,initial}=19\,\mathrm{d}$, the accumulation of accreted \textcolor{black}{non-degenerate} matter on the WD at the beginning of RLOF \textcolor{black}{exceeds the maximum accretion rate for stable H-burning and can lead to a Red-Giant like structure}. In addition, when the accreted matter becomes He-dominated, $\mathrm{log_{10}}(\Dot{M}_2/M_\odot\,\mathrm{yr^{-1}})\ll -6$, which is approximately the boundary of the stable He-burning region. We also check the mass-accumulative efficiency ($\eta$) of nova outbursts from \citet[see their figs~2 and 3]{Wang2018} against our trajectories. They show that for $\mathrm{log_{10}}(\Dot{M}_2/M_\odot\,\mathrm{yr^{-1}})\lesssim-7$, \textcolor{black}{$\eta\lesssim 0.3$ and tends to 0 as $\Dot{M}_2$ decreases further. Thus we can assume that WD hardly retains any accreted matter}. We can conclude from our trajectories that the assumption for fully non-conservative mass transfer is valid for most of our models. Again \textcolor{black}{for} the two most H-exhausted systems with $P_\mathrm{orb,initial}=17\,\mathrm{d}$ and $P_\mathrm{orb,initial}=19\,\mathrm{d}$ \textcolor{black}{the accretor spends a significant amount of time in the  stable H-burning region. This may lead to a} substantial increase in its mass. For these systems, our assumption of non-conservative mass transfer may not be strictly valid (see Section \ref{sec:disc}).

\subsection{Evolved systems in the orbital period distribution of CVs}
\begin{figure}
\includegraphics[width=0.5\textwidth]{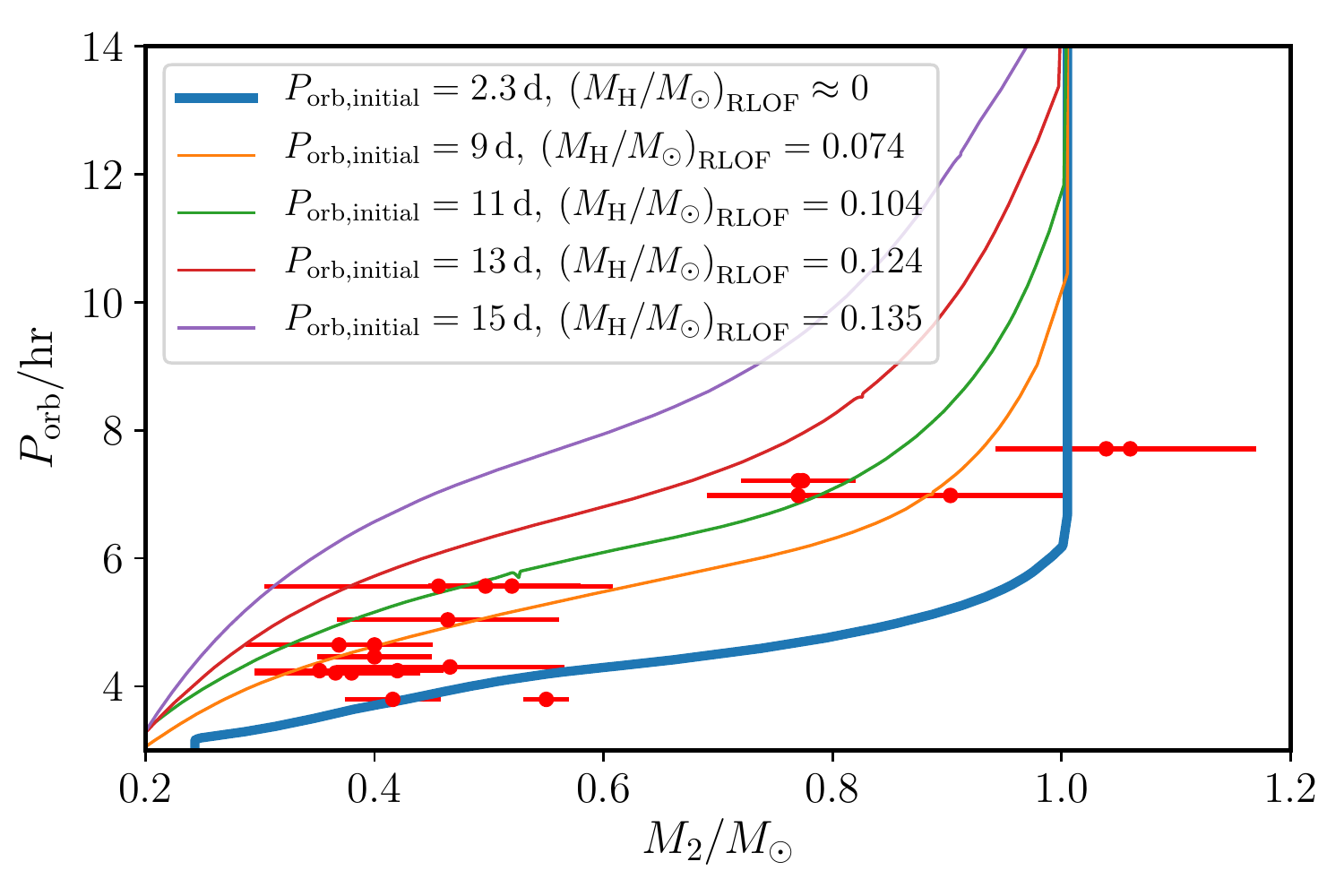}
\caption{Evolutionary tracks of a $M_2=1M_\odot$ and $M_1=1M_\odot$ system in the $(M_2,P_\mathrm{orb})$ plane for different initial orbital periods $P_\mathrm{orb,initial}$ plotted with observed CV data collected by \protect\cite{Ge2015}. \textcolor{black}{The canonical CV discussed earlier (Fig.~\ref{fig:pm_new}) is the thick blue line.} \textcolor{black}{Among the evolved systems, larger values of $P_\mathrm{orb,initial}$ lead to the commencement of RLOF later, leading to more shell H-burning in the subgiant phase of the donor, yielding a larger H-exhausted core. The mass of the H-exhausted core at the beginning of RLOF is also mentioned for each system.}}
\label{fig:pm_hongwei}
\end{figure}

In sections 5.3 and 5.4 of ST we compared \textcolor{black}{the zero-age CV trajectories of} the DD model with the KBP model (see fig.~11) and with observational data of CVs (see figs~12, 13, and 14) taken from \cite{Ge2015}. We found that the long-period CVs fit (fig.~11) and a few systems with $P_\mathrm{orb}\gtrsim5\,\mathrm{hr}$ in fig.~14 did not match very well with our modelled trajectories. We claimed that these systems may be CVs with evolved donors. In Fig.~\ref{fig:pm_hongwei} we plot, with the same observational dataset, the trajectory of systems with $M_2 = 1M_\odot$, $M_1 = 1M_\odot$ and varying $P_\mathrm{orb,initial}$, equally spaced in $P_\mathrm{orb,initial}$ that is likely to form \textcolor{black}{AM CVn stars} with the DD model, taken from this analysis (Section \ref{sec:probanaly}). The system with the smallest $P_\mathrm{orb,initial}$ has the least H-exhausted donor and vice versa.  The system with $P_\mathrm{orb,initial}=9\,\mathrm{d}$ corresponds to the initial conditions with which the system commences RLOF at the end of the MS. We also plot the system in Fig.~\ref{fig:pm_new} which evolves to a canonical CV. It can be seen that donors that commence RLOF when they are mildly H-depleted ($P_\mathrm{orb,initial}=9\,\mathrm{d},11\,\mathrm{d},13\,\mathrm{d}$) match well with systems with $4\,\lesssim P_\mathrm{orb}/\,\mathrm{hr}\lesssim8$ which do not lie on the trajectories of canonical CVs in fig.~11 of ST. 

\begin{figure*}
\centering
\includegraphics[width=0.95\textwidth]{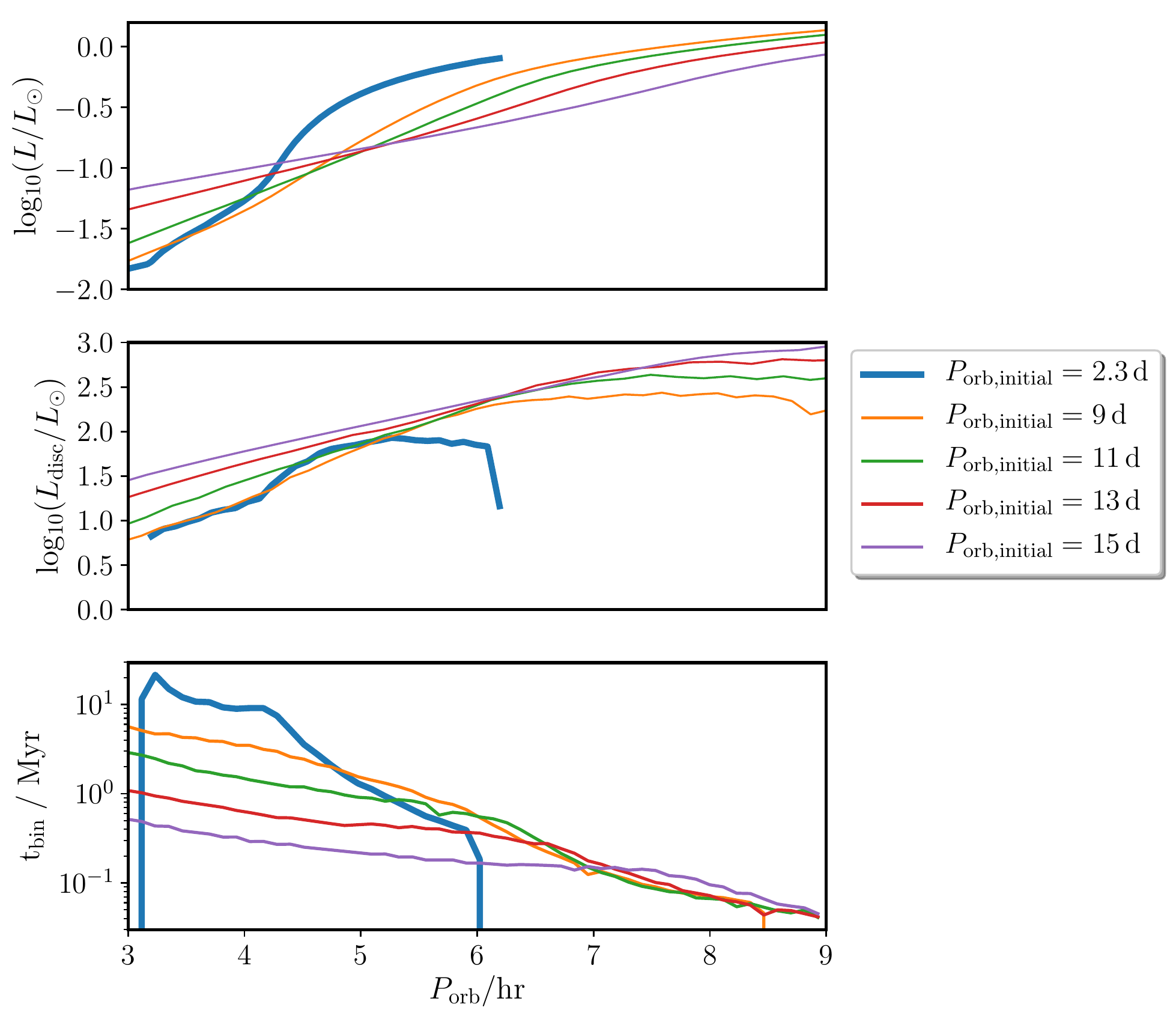}
\caption{\textit{Top and middle}: The evolution of donor luminosity $L_2$ and the disc luminosity $L_\mathrm{disc}$ with $P_\mathrm{orb}$ for the same set of systems as in Fig.~\ref{fig:pm_hongwei}. \textit{Bottom}: The time $t_\mathrm{bin}$ spent by each system in a $P_\mathrm{orb}$ bin for the same set of systems as in Fig.~\ref{fig:pm_hongwei}. \textcolor{black}{The bin size $\mathrm{d}P_\mathrm{orb}\,=\,0.12\,\mathrm{hr}$.}}
\label{fig:PT_hongwei}
\end{figure*}

In the top two panels of Fig.~\ref{fig:PT_hongwei} \textcolor{black}{we show how} the luminosities of the donor $L_2$ and the accretion disc $L_\mathrm{disc}$ evolve with $P_\mathrm{orb}$ after RLOF as a probe of the detection probability of the system. We estimate $L_\mathrm{disc}$ with a simple formula
\begin{center}
\begin{equation}
\label{eq:ldisc}
L_\mathrm{disc} \approx \frac{GM_1\lvert\Dot{M}_2\rvert}{R_1},
\end{equation}
\end{center}
where $M_1,\:R_1,\:\mathrm{and}\:\Dot{M}_2$ are respectively the mass, radius and mass accretion rate of the WD primary. \textcolor{black}{We can see that for a fixed WD primary, $L_\mathrm{disc}$ only depends on $\Dot{M}_2$. Thus, higher accretion rates will lead to a more luminous disc.} We use $R_1\approx 0.008\,R_\odot$ for $M_1\approx 1\,M_\odot$ from \cite{Romero2019}. We can see that for $4.5\,\lesssim P_\mathrm{orb}/\,\mathrm{hr}\lesssim 6\,$, $L_2$ is slightly higher for the canonical CV than that of evolved CVs. However, all donor luminosities become comparable at $P_\mathrm{orb}\approx4\,\mathrm{hr}$. However, it can be seen that $L_\mathrm{disc}$ is likely to dominate the luminosity of the system, wherein evolved systems dominate over the canonical CV at all orbital periods, illustrating that evolved CVs are more likely to be detected at larger orbital periods. In addition, for $P_\mathrm{orb}\gtrsim 6.5\,\mathrm{hr}$, RLOF has not commenced yet for the canonical CV and so for a $1M_\odot$ donor progenitor we expect mass-transferring systems with $P_\mathrm{orb}\gtrsim 6.5\,\mathrm{hr}$ to just be evolved CVs. Less massive donors lead to RLOF \textcolor{black}{beginning} at even shorter $P_\mathrm{orb}$ for canonical CVs. To commence RLOF at $P_\mathrm{orb}\gtrsim 6.5\,\mathrm{hr}$ we need heavier donor progenitors (see the blue and green curves in fig.~A1 of ST). The bottom panel of Fig.~\ref{fig:PT_hongwei} shows the amount of time each system spends in the bin $P_\mathrm{orb}+\mathrm{d}P_\mathrm{orb}$ (\textcolor{black}{with $\mathrm{d}P_\mathrm{orb}\,=\,0.12\,\mathrm{hr}$}) as another probe of the detection probability of the system. The canonical CV spends much more time at $4\,\lesssim P_\mathrm{orb}/\,\mathrm{hr}\lesssim 4.75$ than evolved CVs. However, evolved CVs spend more time at longer $P_\mathrm{orb}$.
Thus, we can conclude that evolved CVs dominate the orbital period distribution at $P_\mathrm{orb}\gtrsim5.5\,\mathrm{hr}$, as \textcolor{black}{indicated by} \cite{Goliasch2015}. 

\subsection{Trajectories of \textcolor{black}{AM CVn stars}}
\label{subsec:AM Cvn}
\begin{figure}
\includegraphics[width=0.5\textwidth]{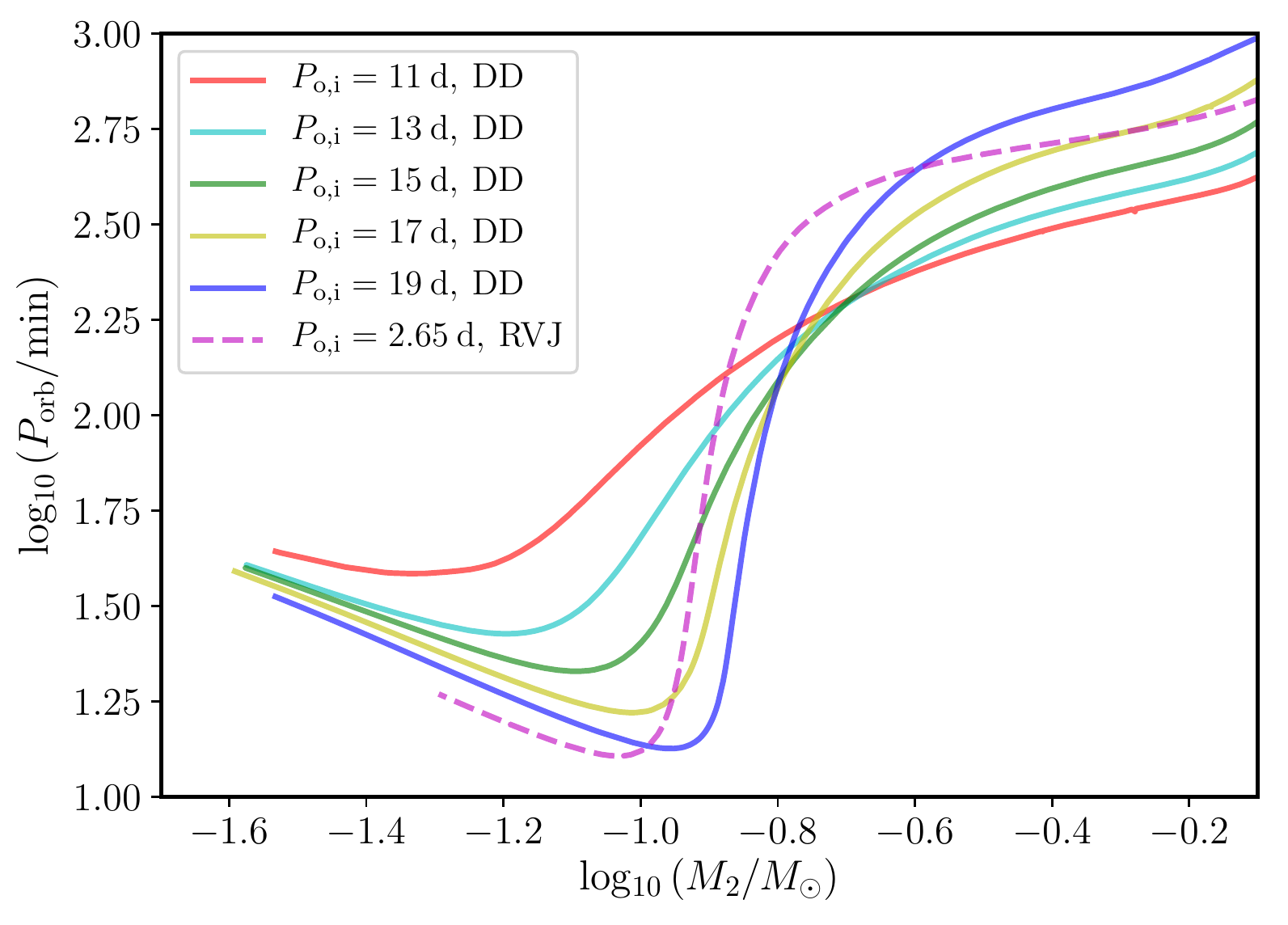}
\caption{The relation of the donor's mass and orbital period in the $(\mathrm{log_{10}}M_2,\mathrm{log_{10}}P_\mathrm{orb})$ plane, \textcolor{black}{for systems with $M_2=1M_\odot$, $M_2=1M_\odot$, and with different $P_\mathrm{orb,initial}$. Increasing $P_\mathrm{orb,initial}$ corresponds to more H-exhausted \textcolor{black}{AM CVn stars}. The solid lines are systems evolved with the DD model and the dashed line is a system evolved with the RVJ model. The systems with $P_\mathrm{orb,initial}=19\,\mathrm{d}$ and $P_\mathrm{orb,initial}=2.65\,\mathrm{d}$ commence RLOF at $t\approx t_\mathrm{BGB}$ with the DD and RVJ model respectively. }}
\label{fig:log-pm}
\end{figure}

\begin{figure}
\includegraphics[width=0.5\textwidth]{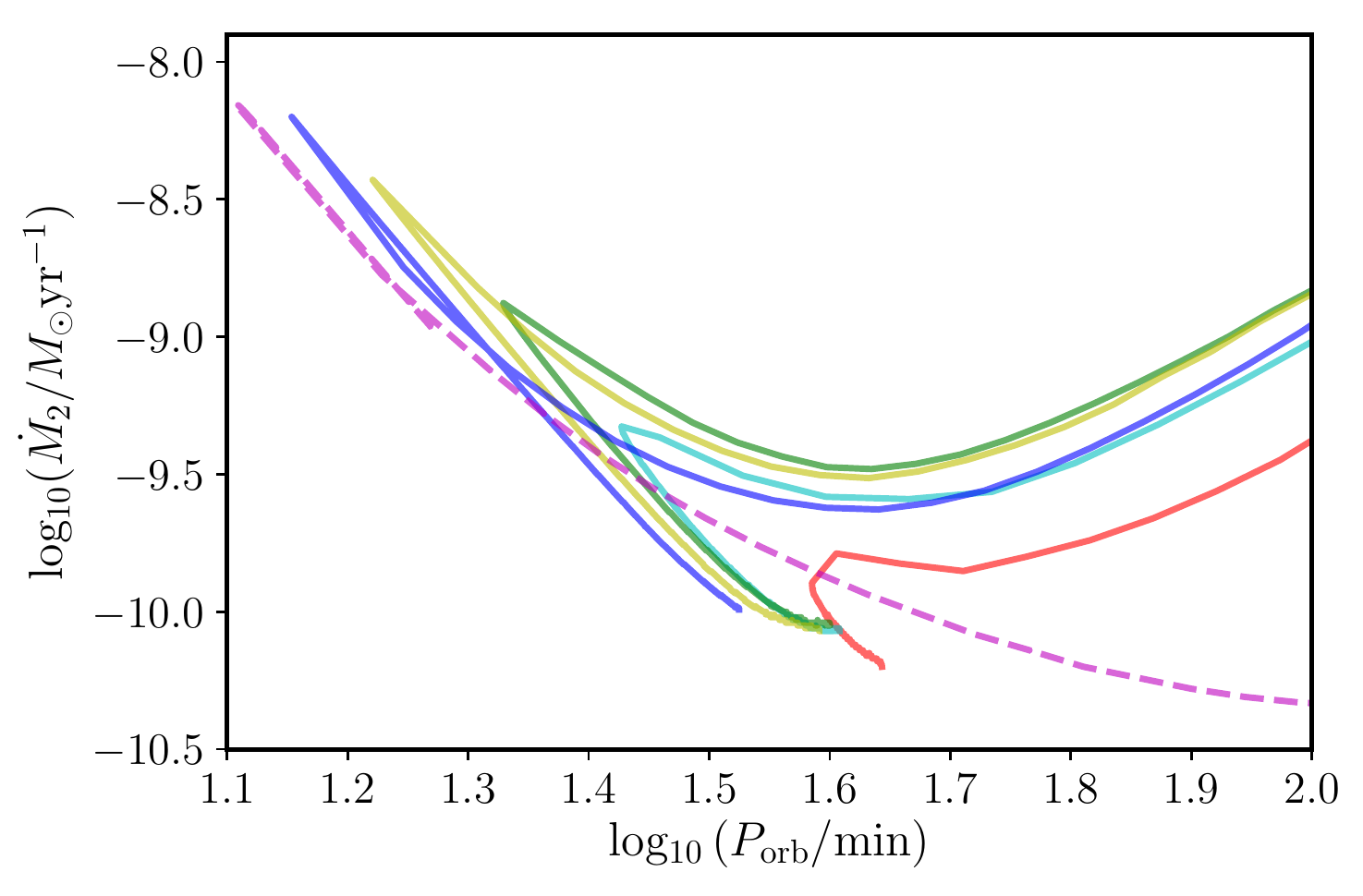}
\caption{The relation of the donor's orbital period and mass-transfer rate in the $(\mathrm{log_{10}}P_\mathrm{orb},\mathrm{log_{10}}\Dot{M}_2)$ plane, for the same set of systems as in Fig.~\ref{fig:log-pm}.}
\label{fig:log-pmdot}
\end{figure}

We make detailed trajectories of evolved CVs, and analyse them beyond their orbital period minima, when they become ultracompact CVs or \textcolor{black}{AM CVn stars} through the Evolved CV channel. This is shown in Figs~\ref{fig:log-pm} and \ref{fig:log-pmdot} for the DD model (solid curves) and the RVJ model (dashed curves) of $\mathrm{AML_{MB}}$. The system with $P_\mathrm{orb,initial}=19\,\mathrm{d}$ corresponds to the initial conditions with which the system commences RLOF at the base of the giant branch in the DD model. We also plot a system with $P_\mathrm{orb,initial}=2.65\,\mathrm{d}$ which corresponds to the initial conditions from which the system commences RLOF at $t\approx t_\mathrm{BGB}$ with the RVJ model. \textcolor{black}{In Fig.~\ref{fig:log-pm} we can see that for the DD model, systems with larger $P_\mathrm{orb,initial}$ have more compact donors (tighter orbit) for a given $M_2$ after their period minima. We mentioned in connection with Fig.~\ref{fig:pm_hongwei}, that systems with larger $P_\mathrm{orb,initial}$ have larger H-depleted cores at the beginning of RLOF, thus yielding AM CVn systems with tighter orbits after their period minima. This trend is seen in general for the RVJ model as well. It is interesting to compare the most H-exhausted trajectory of the DD model with the RVJ model. Owing to the qualitatively different prescription of the AML implementation (see Section~\ref{sec:probanaly}), there are differences in the behaviour of the system at the same donor mass and a direct comparison of the two is not straightforward. However, some conclusions can still be drawn. We have seen that at the commencement of RLOF, the RVJ model is much more compact than the DD model, and the evolution just after the beginning of RLOF is much faster for the DD model compared to the RVJ model. At $M_2\approx 0.3\,M_\odot$, the braking time-scale of the DD model exceeds that of the RVJ model, and owing to this $P_\mathrm{orb}$ for the RVJ model is larger than that of the DD model for a given donor mass for $M_2\lesssim 0.3\,M_\odot$. This is also the reason why the RVJ trajectory attains its period minimum for a larger donor mass than the DD model. However, after the period minimum, the RVJ model leads to a much more compact donor compared to its DD counterpart because the driving AML mechanism becomes only $\mathrm{AML_{GR}}$ for the RVJ model, whereas the DD model incorporates $\mathrm{AML_{GR}}+\mathrm{AML_{DD}}$  with increasing contribution from the enhanced convective dynamo as the orbit widens and the donor becomes increasingly convective (see ST, and \citealt{2023MNRAS.519.2567S} for details on the contribution of $\mathrm{AML_{DD}}$ for AM CVn stars). Very importantly, the time taken for this system to evolve with the RVJ model is greater than $12\,\mathrm{Gyr}$. So in reality this system would not form within the Galactic age. We plot its trajectory anyway to compare and contrast the two models' characteristic parameters. We note that the current formulation of the DD model can also yield more compact donors than discussed here if RLOF commences at times later than $t_\mathrm{BGB}$. However, simulations with the RVJ model suggest that systems where the donor is in its giant branch respond to mass transfer by widening their orbit, thereby transitioning from ultracompact binaries to wide binaries \citep{Nelson2004}. We do not attempt to model these systems with the DD model in this work because the current formulation of the DD model fails to produce wide binaries and still yields ultracompact systems (Section~\ref{sec:bif}). } It can also be seen that, beyond their period minima, each trajectory tends to follow a linear relation in the $(\mathrm{log_{10}}M_2,\mathrm{log_{10}}R_2)$ and $(\mathrm{log_{10}}M_2,\mathrm{log_{10}}P_\mathrm{orb})$ plane. We fit these trajectories and extrapolate our curves in Section \ref{subsubsec:fit}. Finally, Fig.~\ref{fig:log-pmdot} shows the mass-transfer rate $\dot{M}_2$ with $P_\mathrm{orb}$. It can be seen that the RVJ model has significantly smaller $\dot{M}_2$ at longer periods than the DD model. However the \textcolor{black}{two models} tend to become qualitatively similar as the trajectories approach their period minimum.


\subsection{Comparison with observations}
\label{subs:obs}
We compare our modelled systems to observations. We begin by comparing the observed parameters of \textcolor{black}{AM CVn stars} to our modelled trajectories and then model the mass fractions of H and CNO-processed elements. We limit our analysis to a set of systems with initial $M_2/M_\odot=\{1,1.2\}$ and $M_1=1M_\odot$, \textcolor{black}{because the most important factor governing the evolution of our trajectories is the extent of H-exhaustion in the donor, which can be well modelled just by increasing or decreasing $P_\mathrm{orb,initial}$ (Section~\ref{subsec:AM Cvn}). We still discuss the influence of varying $M_1$ and $M_2$ on our trajectories for completeness.} \textcolor{black}{Increasing $M_1$ lowers $\Dot{M}_2$ and increases the evolution time-scale. As a result more H-burning can occur during the evolution of the system, which in turn makes the donor more H-exhausted for the same initial $M_2$ and $P_\mathrm{orb,initial}$. The dependence of the observed parameters on the initial $M_2$ in our set of models is weak, because the final trajectory is only dependent on the compactness of the donor, which (at least for the DD model) depends on the stage between the donor's TMS and BGB phase when RLOF commences. In general, heavier secondaries will have a more massive H-depleted core which will yield a more compact donor for a given $M_2$ close to their period minimum. The CNO abundances are also dependent on $M_2$ owing to the dependence of CNO equilibrium on the mass of the star, as discussed in Section~\ref{sec:cno+he}. }

\subsubsection{Analysis of observed parameters}

\begin{figure*}
\centering
\includegraphics[width=0.95\textwidth]{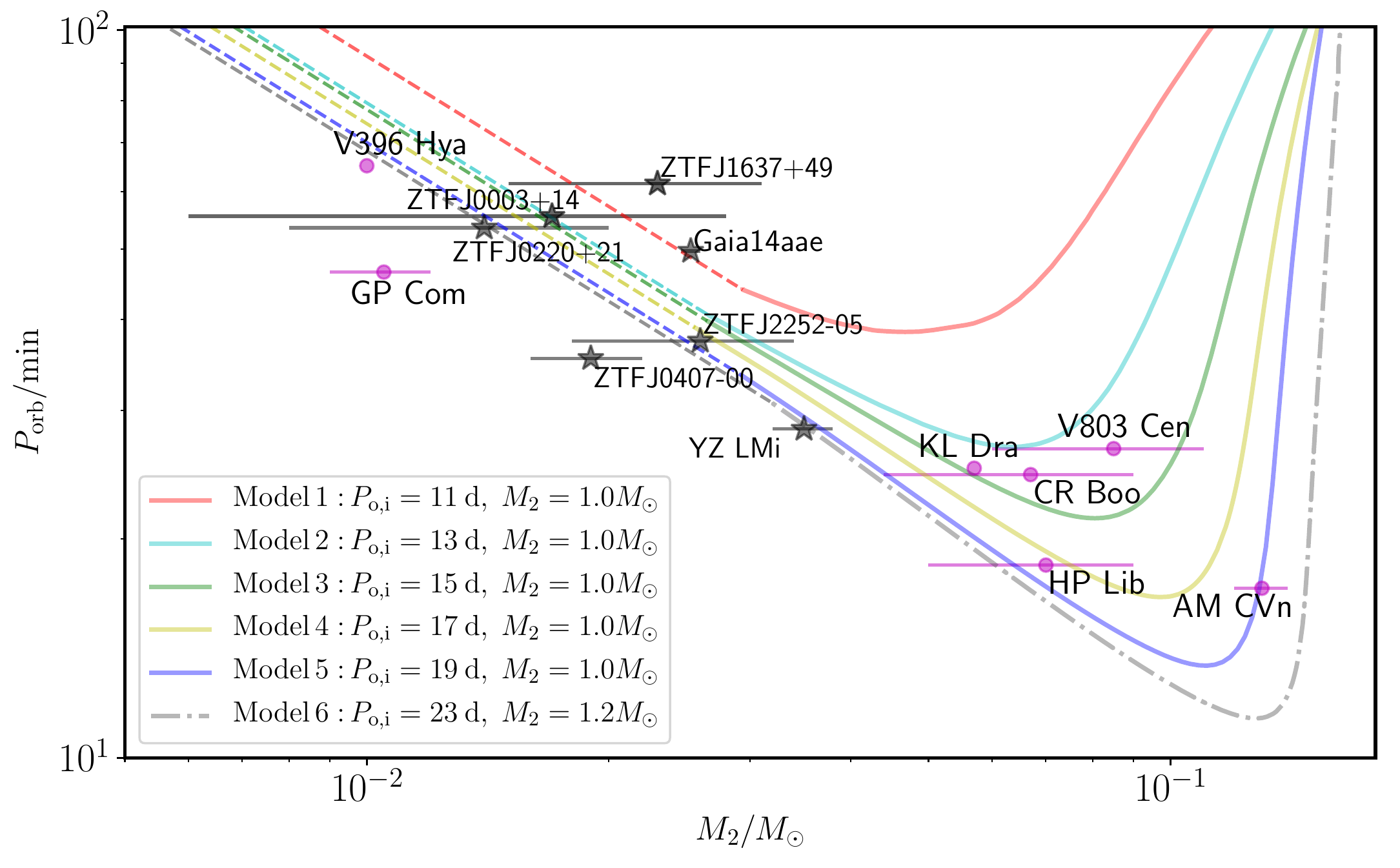}
\caption{The relation of the donor's orbital period and mass in the $(M_2,P_\mathrm{orb})$ plane, for systems with $M_1=1M_\odot$ and different $P_\mathrm{orb,initial}$ ($P_\mathrm{o,i}$). Increasing $P_\mathrm{orb,initial}$ corresponds to more H-exhausted AM CVn stars. The systems with $P_\mathrm{orb,initial}=19\,\mathrm{d}$ and $P_\mathrm{orb,initial}=23\,\mathrm{d}$ commence RLOF at $t\approx t_\mathrm{BGB}$ for different donor masses. The dashed section in each trajectory is a power law fit of the form $P_\mathrm{orb} \propto (M_2/M_{2,P_\mathrm{orb,min}})^\beta$, where $\beta=-0.6860$. The points in magenta are systems from \protect\cite{Solheim2010}, while the stars in black are systems described by \protect\cite{2022MNRAS.512.5440V}, \protect\cite{Green2018} and \protect\cite{Copperwheat2010}.}
\label{fig:pm-solvan}
\end{figure*}

\begin{figure*}
\centering
\includegraphics[width=0.95\textwidth]{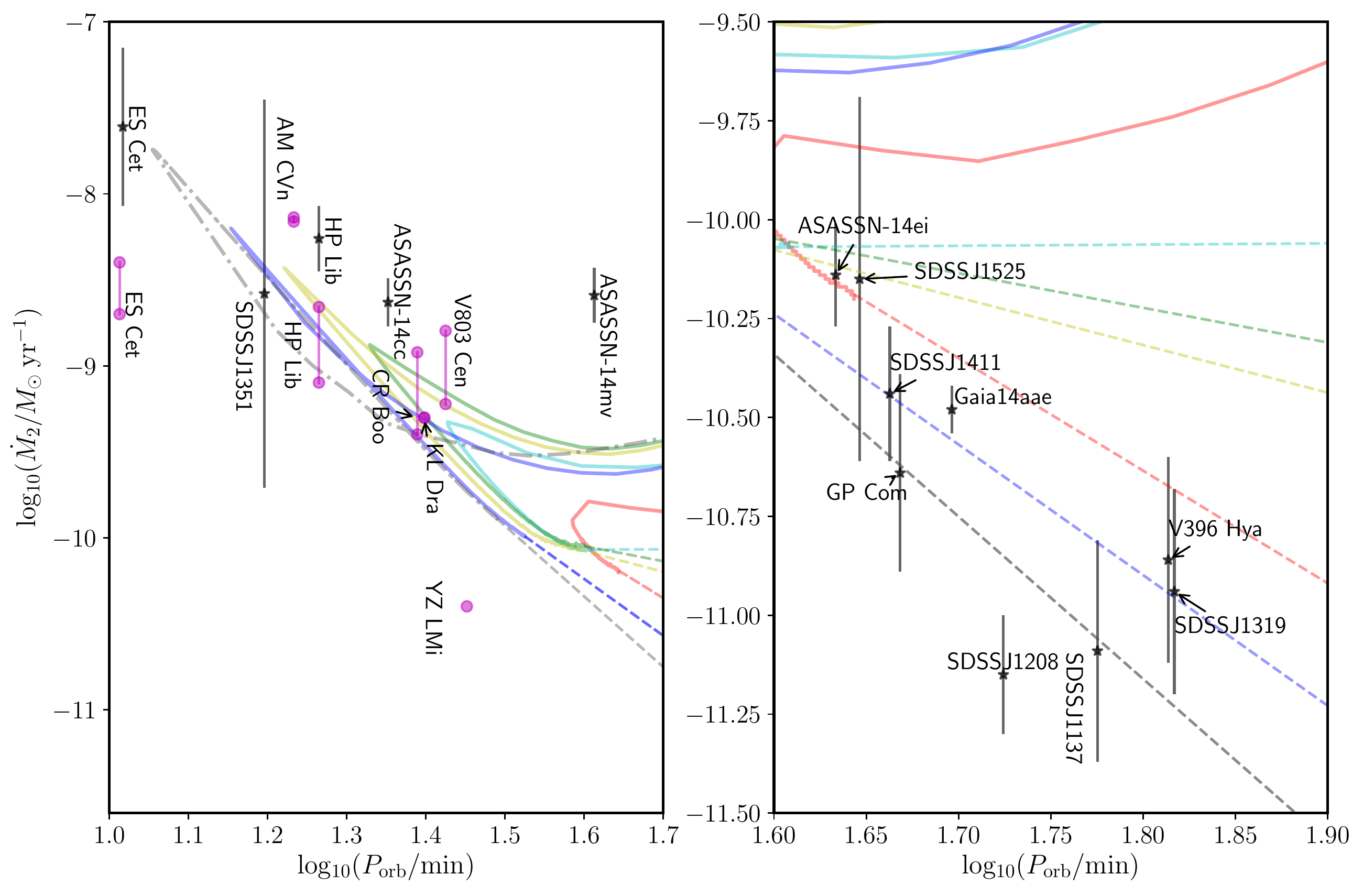}
\caption{The relation of the donor's orbital period and mass-transfer rate in the $(\mathrm{log_{10}}P_\mathrm{orb},\mathrm{log_{10}}\Dot{M}_2)$ plane for the same set of trajectories as in Fig.~\ref{fig:pm-solvan}. The dashed section in each trajectory is a power-law fit of the form $\dot{M}_2\propto P_\mathrm{orb}^\gamma$. The points in magenta are systems reported by \protect\cite{Solheim2010}, while the stars in black are the systems described by \protect\cite{2018A&A...620A.141R}. Except for the eclipsing system (Gaia14aae), they use \textcolor{black}{$M_1 = 0.8\pm0.1M_\odot$} and an inclination angle such that $\mathrm{cos}\:i=0.5$ to obtain the mass-transfer rates.}
\label{fig:mdotP-solram}
\end{figure*}




\label{subsubsec:fit}

We first look at the donor mass--orbital period relationship of the system and compare it with five eclipsing \textcolor{black}{AM CVn stars} with well-defined $P_\mathrm{orb}$ and $M_2$  \citep{2022MNRAS.512.5440V}, Gaia14aae  \citep{Green2018} and YZ LMi \citep{Copperwheat2010}. In addition, we also compare our results to systems reported by \cite{Solheim2010}. \textcolor{black}{We compare their donor parameters with trajectories evolved with the DD model, spaced in $P_\mathrm{orb}$ such that we obtain models ranging from the least H-exhausted system to the most H-exhausted system.} We see that most of these systems lie in a region of the $(M_2,\,P_\mathrm{orb})$ plane where our code cannot be used to model them\footnote{The code fails for $M_2\lesssim0.03M_\odot$ for the DD model \textcolor{black}{because the surface of the star becomes degenerate.}}. So we extrapolate to evolve to smaller $M_2$. As seen in Fig.~\ref{fig:log-pm}, after a minimum in $\mathrm{log\,}P_\mathrm{orb}$, the trajectories follow a linear relation in log space, tending towards a common slope. This is visible in the most H-exhausted systems where $P_{\mathrm{orb},\mathrm{min}}$ is at a larger $M_2$. Thus we extrapolate our models with a power law function of the form

\begin{center}
\begin{equation}
\label{eq:beta}
P_\mathrm{orb}(M_2) = P_\mathrm{orb}(M_{2,P_\mathrm{min}})\left(\frac{M_2}{M_{2,P_\mathrm{min}}}\right)^\beta,
\end{equation}
\end{center}
where $M_{2,P_\mathrm{min}}$ is $M_2$ when $P_\mathrm{orb}$ attains a minimum. We approximate $\beta$ by averaging over the slopes of the three most H-exhausted systems for the $M_2=1M_\odot$ star, that is, systems with $P_\mathrm{orb,initial}\,/\mathrm{d}=\{15,17,19\}$, which gives us $\beta=-0.6860$. Our result is shown in Fig.~\ref{fig:pm-solvan}. \textcolor{black}{From the mass--orbital period data we find good agreement of our models with systems such as Gaia14aae (Model 1), ZTFJ1637+49 (Model 1), ZTFJ0003+14 (all models), ZTFJ0220+21 (Model 2, 3, 4, 5 and 6), ZTFJ2252-05 (Model 2, 3, 4, 5 and 6), YZ LMi (Model 5 and 6), V396 Hya (mild agreement with Model 5 and 6), KL Dra (Model 3), CR Boo (Model 3, 4 and 5), V803 Cen (Model 2 and 3), HP Lib (Model 4, 5 and 6) and AM CVn (Model 5).}

We also plot the evolution of $\Dot{M}_2$ with $P_\mathrm{orb}$ in Fig.~\ref{fig:mdotP-solram}. We do not see any clear trend between different systems, so we simply extrapolate each system separately by fitting a power law function of the form 
\begin{center}
\begin{equation}
\label{eq:gamma}
\Dot{M_2} \propto P_\mathrm{orb}^\gamma
\end{equation}
\end{center}
to the last 200 models of each trajectory.  We note that these extrapolated trajectories are less reliable owing to the absence of a clear trend. The mass transfer also depends on the mass of the accretor $M_1$, with smaller $M_1$ leading to higher mass-transfer rates. We also note that different mass-transfer rates have been reported by different groups for the same system (for instance HP Lib and ES Cet in Fig.~\ref{fig:mdotP-solram}). So we focus more on $M_2$ and $P_\mathrm{orb}$ than $\Dot{M}_2$. \textcolor{black}{We find agreement of our models with observations, particularly systems from Fig.\ref{fig:pm-solvan} such as V396 Hya (Model 1 and 5), KL Dra (Model 3 and 5), CR Boo (Model 3, 4, 5 and 6), V803 Cen (Model 3 and 4), GP Com (Model 5 and 6) and HP Lib (Model 4, 5 and 6) where the mass-transfer rate agrees with data of \cite{Solheim2010}. We note here that although GP~Com matches well with our trajectories in Fig.~\ref{fig:mdotP-solram}, owing to poorer constraints on the mass-transfer rates we rely more on its $M_2-P_\mathrm{orb}$ relation. The $M_2$ and $P_\mathrm{orb}$ values of GP~Com do not match with any of our trajectories in Fig.\ref{fig:pm-solvan}. Its much tighter orbit for its donor mass suggests a degenerate donor \citep{Roelofs2007} and rules out the Evolved CV formation channel.} Interestingly, the suggested donor star by \cite{Solheim2010} for many of these systems is either a He-star or a WD with the exception of Gaia14aae, \textcolor{black}{of which the progenitor has been suggested to be an evolved H-star \citep{Campbell2015,Green2018}, although recent results of \cite{2023MNRAS.519.2567S} show that Gaia14aae and ZTFJ1637+49 likely formed through the He-star channel}. The results of \cite{Roelofs2007} also suggest a semi-degenerate donor for AM CVn, HP Lib, CR Boo, and V803 Cen.

\subsubsection{Analysis of abundances}
\label{sec:cno+he}

\begin{figure*}
\centering
\includegraphics[width=0.95\textwidth]{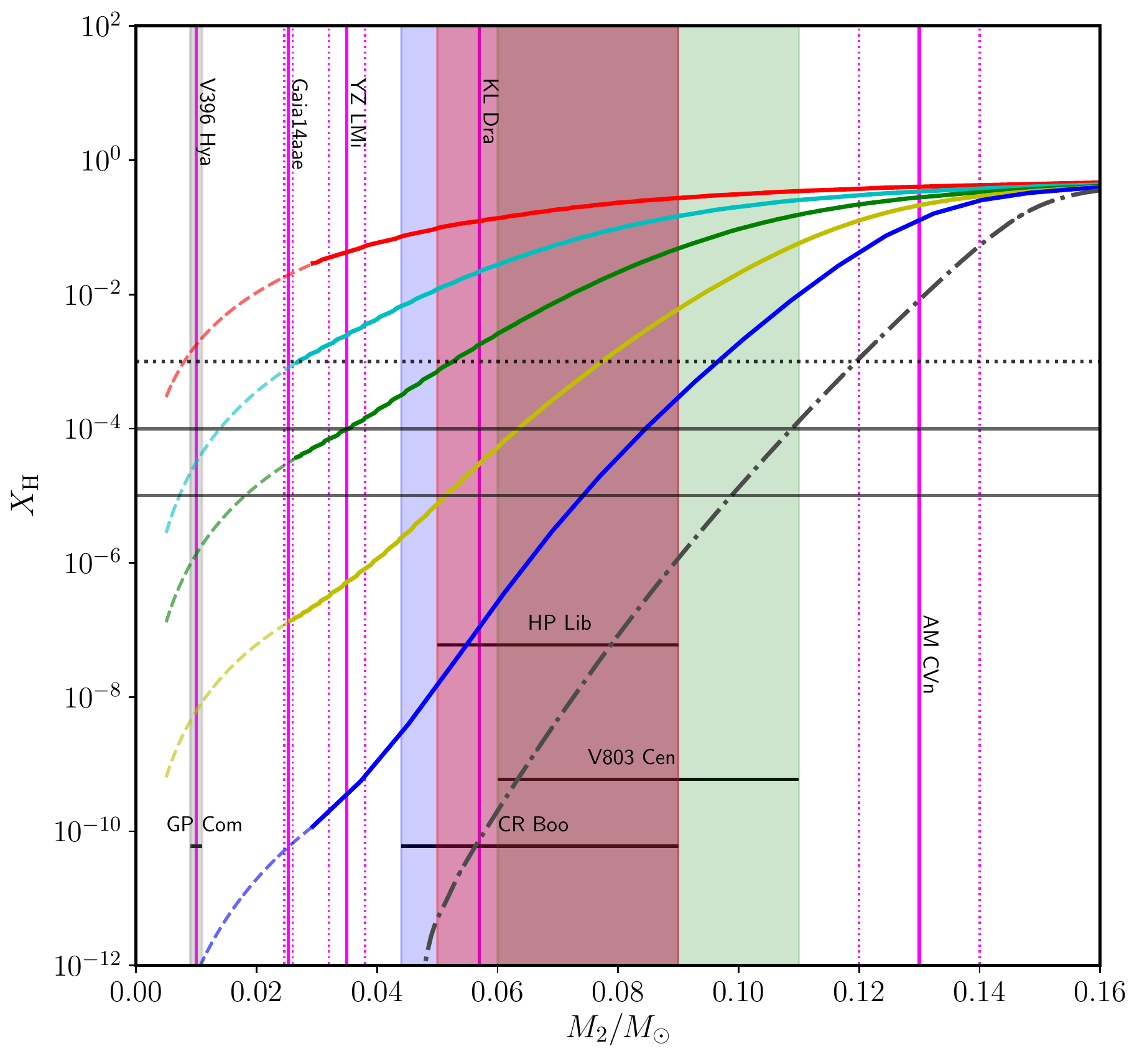}
\caption{The surface H abundance $X_\mathrm{H}$ of the donor with $M_2$ for the same modelled systems as in Fig.~\ref{fig:pm-solvan}. The vertical solid lines and the associated dotted lines in magenta are the donor masses and the error bars of observed systems obtained by \protect\cite{Solheim2010} and \protect\citet[and the references therein]{2022MNRAS.512.5440V}. The shaded regions are donors that have an inferred mass range. V083 Cen (green) has $M_2\in[0.06,\,0.11]M_\odot$, HP Lib (red) has $M_2\in[0.05,\,0.09]M_\odot$, CR Boo (blue) has $M_2\in[0.044,\,0.09]M_\odot$, and GP Com (grey) has $M_2\in[0.009,\,0.012]M_\odot$. Horizontal lines in black denote $X_\mathrm{H}=10^{-4}$ and $10^{-5}$ while the dotted one denotes $X_\mathrm{H}=10^{-3}$.}
\label{fig:x}
\end{figure*}

\begin{figure*}
\centering
\includegraphics[width=0.95\textwidth]{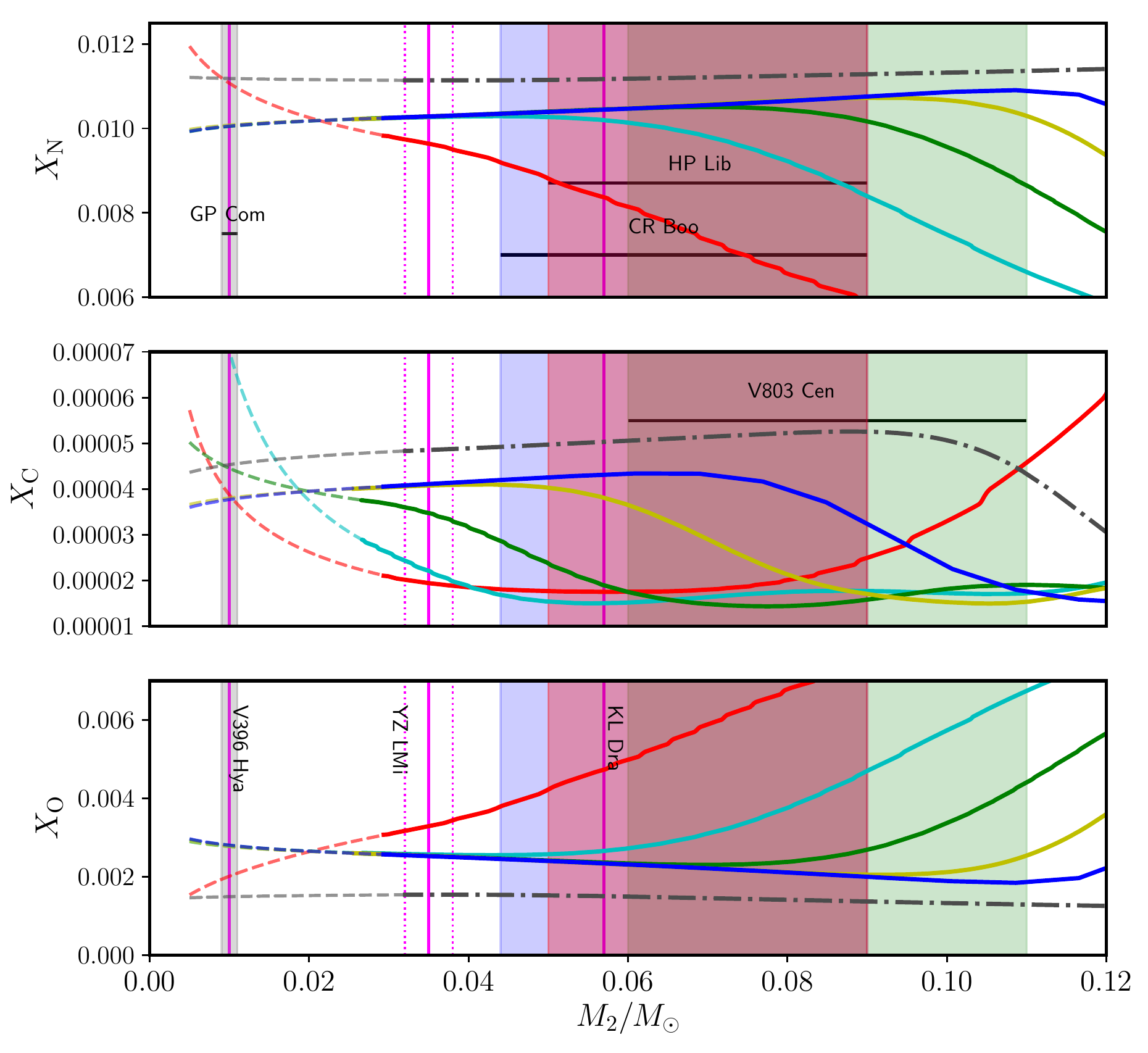}
\caption{The surface abundances of N, C, and O of a donor of mass $M_2$ for the same modelled systems as in Fig.~\ref{fig:pm-solvan}. The vertical solid lines and the associated dotted lines in magenta are the donor masses and the error bars of observed systems obtained by \protect\cite{Solheim2010} and \protect\citet[and the references therein]{2022MNRAS.512.5440V} as in Fig.~\ref{fig:x} but omitting Gaia14aae and AM CVn.}
\label{fig:xother}
\end{figure*}

We can also model the H, C, O, and N abundances at the surface of the donor and thence the accretion disc. We first look at the mass fraction of H, $X_\mathrm{H}$, which has been used to distinguish between the Evolved CV channel and the other channels. It has been argued that $X_\mathrm{H}\gtrsim10^{-5}$ would trigger detectable Balmer emission \citep{Green2018}. \cite{Green2019} states that $X_\mathrm{H}\approx 10^{-4}$ cannot be ruled out either. However, the work of \cite{2009Nagel} shows that in high-state systems (high-mass transfer rate, typically $\Dot{M}_2\approx 10^{-8}M_\odot\,\mathrm{yr^{-1}}$), $X_\mathrm{H}\approx 10^{-3}$ is needed to obtain detectable Balmer lines. \textcolor{black}{So we use the upper bound of $10^{-4}$ and a lower bound of $10^{-5}$ on the minimum $X_\mathrm{H}$ for detection for systems with $\Dot{M}_2 \lesssim 10^{-10} M_\odot\,\mathrm{yr^{-1}}$ and a minimum $X_\mathrm{H}\approx 10^{-3}$ for detection for systems with $\Dot{M}_2 \gtrsim 10^{-10} M_\odot\,\mathrm{yr^{-1}}$. Any AM CVn with $X_\mathrm{H}$ below these limits should have no trace of H in its spectrum making the system indistinguishable, in terms of the absence of H, from those evolved through the WD channel or the He-star channel.} We classify systems as high-state and low-state by looking at Fig.~\ref{fig:mdotP-solram}. We see that among the systems discussed in Fig.~\ref{fig:pm-solvan}:
\begin{enumerate}
    \item the high-state systems are AM CVn, HP Lib, V803 Cen, CR Boo, KL Dra, requiring a minimum $X_\mathrm{H}\approx10^{-3}$ for detection.
    \item the low-state systems are GP Com, Gaia14aae, V396 Hya and YZ LMi, requiring a minimum $X_\mathrm{H}\approx 10^{-4}$ for detection.
\end{enumerate}
Our result is shown in Fig.~\ref{fig:x} in the $(X_\mathrm{H},M_2)$ plane, where we also plot the observed donor masses of the systems in which we are interested. We extrapolate the trajectories with a simple power law of the form $X_\mathrm{H}\propto M_2^\delta$, where we obtain $\delta$ for each trajectory by fitting the last 200 models.

Inferring $X_\mathrm{H}$ for a system requires using the results of Section~\ref{subsubsec:fit} and Fig.~\ref{fig:x}. For Gaia14aae we obtain good agreement of observed data with the mass--orbital period trajectory of Model 1. So we infer its $X_\mathrm{H}$ by finding the intersection of the vertical line in magenta corresponding to the donor mass of Gaia14aae with this model trajectory in Fig.~\ref{fig:x}. This gives us $X_\mathrm{H} \approx 10^{-2}$. \textcolor{black}{For V803 Cen, we obtain good agreement of observed data with the mass--orbital period trajectory of Model 2 and 3, wherein Model 2 matches if its actual donor mass is at the lower-end of the interval, while Model 3 matches if its actual donor mass is at the higher end of the interval. We look at the intersection of these trajectories with the green vertical shaded region corresponding to the observed donor mass range of V803 Cen in Fig.~\ref{fig:x}. We see that the inferred $X_\mathrm{H}$ is the intersection of the lower bound of the shaded region with Model 2 or the intersection of the higher bound of the shaded region with Model 3, both of which give $X_\mathrm{H} > 10^{-2}$.} Similarly, we find that for YZ LMi $X_\mathrm{H} \lesssim 10^{-10}$, for V396 Hya $X_\mathrm{H} \approx 10^{-12}$, for KL Dra $X_\mathrm{H} > 10^{-3}$, for CR Boo $10^{-9}\lesssim X_\mathrm{H} \lesssim 5\times10^{-2}$, for HP Lib $10^{-11}\lesssim X_\mathrm{H} \lesssim 5\times10^{-3}$, and for AM CVn $X_\mathrm{H}\approx5\times10^{-2}$. Based on these mass-fraction estimates, and the fact that the spectra of all these systems are H-deficient meaning that depending on the mass-transfer rate of the system their inferred $X_\mathrm{H}$ should be below the minimum $X_\mathrm{H}$ for detection, \textcolor{black}{we can safely rule out the Evolved CV formation channel for AM CVn, Gaia14aae, and V803~Cen. KL~Dra can also be ruled out if we assume that $X_\mathrm{H}=10^{-3}$ is a sharp cut-off between detection and non-detection. We have also ruled out GP~Com owing to its donor being more compact than our most H-exhausted trajectories. CR~Boo and HP~Lib have Evolved CV channel as a viable possibility if we assume that their actual donor masses lie on the lower-end of the observed interval. YZ~LMi and V396~Hya match with the Evolved CV channel trajectories comfortably. We still track their abundances for completeness.}

\textcolor{black}{We briefly revert to the discussion in Section~\ref{subs:obs} about varying the initial $M_1$ and $M_2$ in our trajectories and its effect on the comparison with observations. For instance, Model 3 matches well with the mass and orbital period of KL~Dra, but it yields an inferred H-abundance that ought to be detectable according to our selected lower limit of $10^{-3}$. For a larger $M_1$ or $M_2$, the system will be more H-depleted and can yield an inferred $X_\mathrm{H}<10^{-3}$. However, due to being more compact such a trajectory will not match well with the mass and orbital period of KL~Dra. Thus, the requirement of a trajectory to satisfy both, the observed parameter constraint, as well as the abundance constraint, reduces the sensitivity of our results to changes in $M_1$ and $M_2$ with which we begin our evolution.}  

The C, N, and O abundances of the systems above are shown in Fig.~\ref{fig:xother}. We see that the most H-exhausted donors saturate at surface abundances of $X_\mathrm{N}\approx0.01$, $X_\mathrm{C}\approx4\times10^{-5}$ and $X_\mathrm{O}\approx0.002$. \textcolor{black}{For the system with $M_2=1.2\,M_\odot$, the CNO abundance consists of enhanced N and C, and lowered O relative to its $M_2=1\,M_\odot$ counterpart. This can be attributed to the fact that a heavier star undergoing shell H-burning in the subgiant phase (wherein the CNO cycle also operates) attains CNO equilibrium at a different temperature, thus leading to different abundances of C, N and O.} HP Lib and CR Boo have reported $X_\mathrm{N}$ from X-rays at $0.02$ and $0.015$ respectively (see Table~1 of \citealt{Nelemans2010} and the references therein). These are quite close to what we find for typical models, \textcolor{black}{and a better match to the observed $X_\mathrm{N}$ values can be obtained with a more massive initial donor mass.}

\section{Discussion}
\label{sec:disc}

\textcolor{black}{In this section we consider some of our assumptions in more detail, and comment on the shortcomings of our analysis and their possible consequences.}
\subsection{Non-conservative mass transfer and novae}
As discussed in Section \ref{sec:model}, our assumption of purely non-conservative mass transfer may not be valid during the initial phase of RLOF for the most H-exhausted systems with the DD model (Fig.~\ref{fig:mdot-t}). \textcolor{black}{A high mass-transfer rate leads to the accumulation of mass on the WD and the deposition of H-rich matter in massive accretors may lead to a type Ia supernova by the single-degenerate channel if the resulting mass exceeds $1.38\,M_\odot$ (\citealt{1973ApJ...186.1007W}; also see \citealt{Wang2018} for a thorough review of accreting WDs and type Ia supernovae). As a consequence, some systems with massive accretors that commence RLOF around the BGB of the donor do not evolve to form \textcolor{black}{AM~CVn stars}. This reduces the parameter space of expected AM CVn progenitors with the DD model (Section \ref{sec:probanaly}).}
    
\subsection{Distribution of orbital separation and orbital period} 
\textcolor{black}{In our birthrate analysis (Section \ref{sec:probanaly}) we use two distribution functions for the donor mass, one being the Salpeter IMF and the other being a uniform distribution, with a flat distribution in $\mathrm{log}\,a$}. An analysis of the distribution of orbital semi-major axes in post-common-envelope WD systems as a function of $M_1$ and $M_2$ by \citet[see figs~9e and 9f]{Kruckow2021} found that most donors with $M_2$ of our interest have $a\approx 10R_\odot$ while relatively fewer systems have larger $a$ and even fewer have $a\lesssim5R_\odot$. So our assumption of a flat distribution of $\mathrm{log}\,a$ may not be strictly valid. \textcolor{black}{As a consequence, our analysis in Section \ref{sec:probanaly} can overestimate the expected number of \textcolor{black}{AM CVn stars}.}
    
\subsection{Modelling heavy donors} 
We work mainly with the DD model as described by ST and \cite{Zangrilli1997}. This only works efficiently when the donor has a convective envelope and a radiative core. Because donor stars with $M_2\gtrsim1.4M_\odot$ develop a radiative envelope with a convective core, we expect the DD model of $\mathrm{AML_{MB}}$ to be altered. This has neither been studied by ST nor here. So we limit our analysis to donors with $M_2\leq1.3M_\odot$ as discussed in Section \ref{sec:probanaly}. Thus these analyses could underestimate the parameter space of AM CVn progenitors because a considerable number of heavier donors might also evolve to form \textcolor{black}{AM CVn stars} through the Evolved CV channel. This could increase the number of expected \textcolor{black}{AM CVn stars} calculated at the end of Section \ref{sec:probanaly}.

\subsection{The bifurcation limit}
\label{sec:bif}

\begin{figure}
\includegraphics[width=0.5\textwidth]{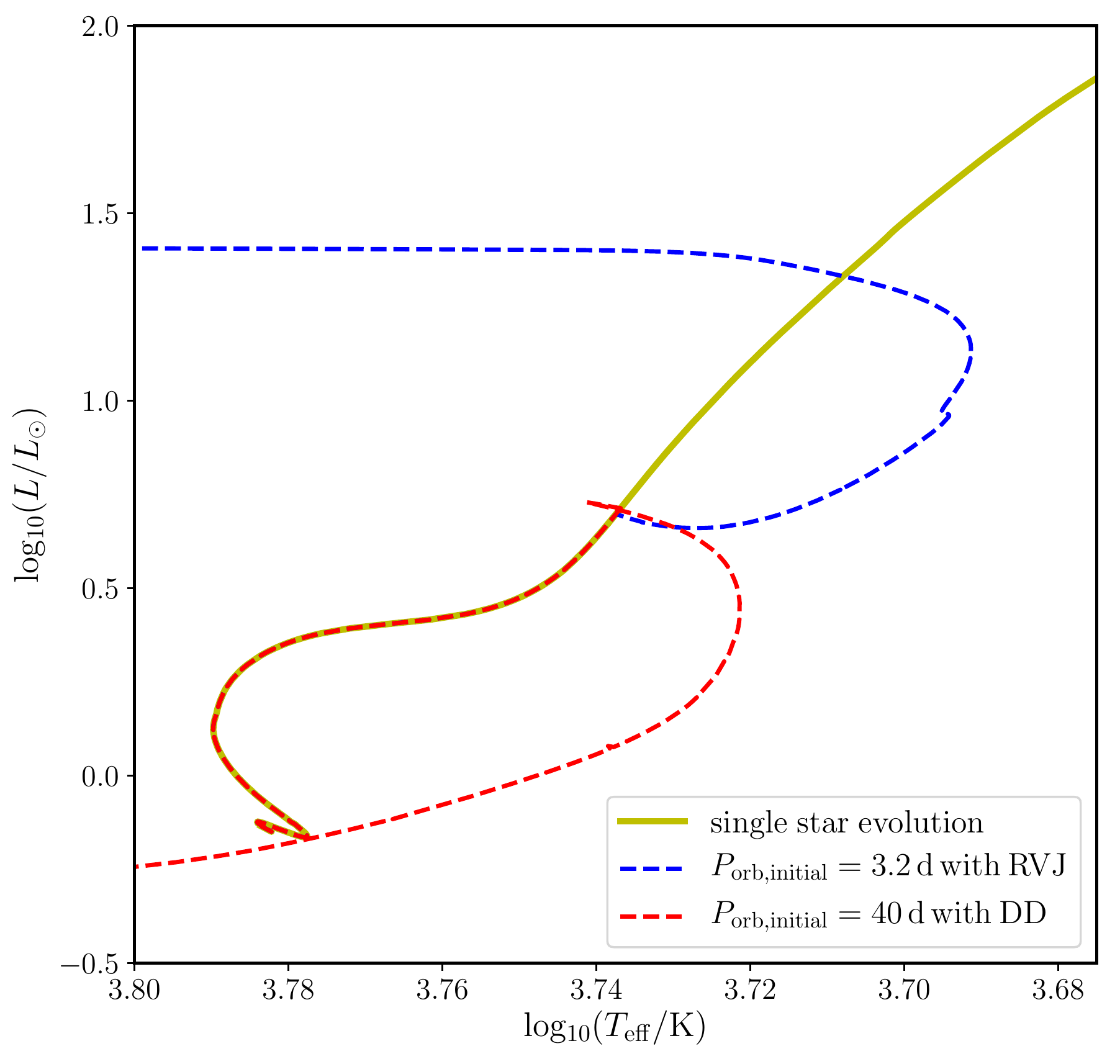}
\caption{The HR diagram of a $1M_\odot$ donor star with $M_1=1M_\odot$ and $P_\mathrm{orb,initial}$ such that RLOF commences at the same evolutionary stage after $t_\mathrm{BGB}$ but with differing $\mathrm{AML_{MB}}$, along with the HR diagram of an isolated $1M_\odot$ star.}
\label{fig:hr-giant}
\end{figure}

\begin{figure}
\includegraphics[width=0.5\textwidth]{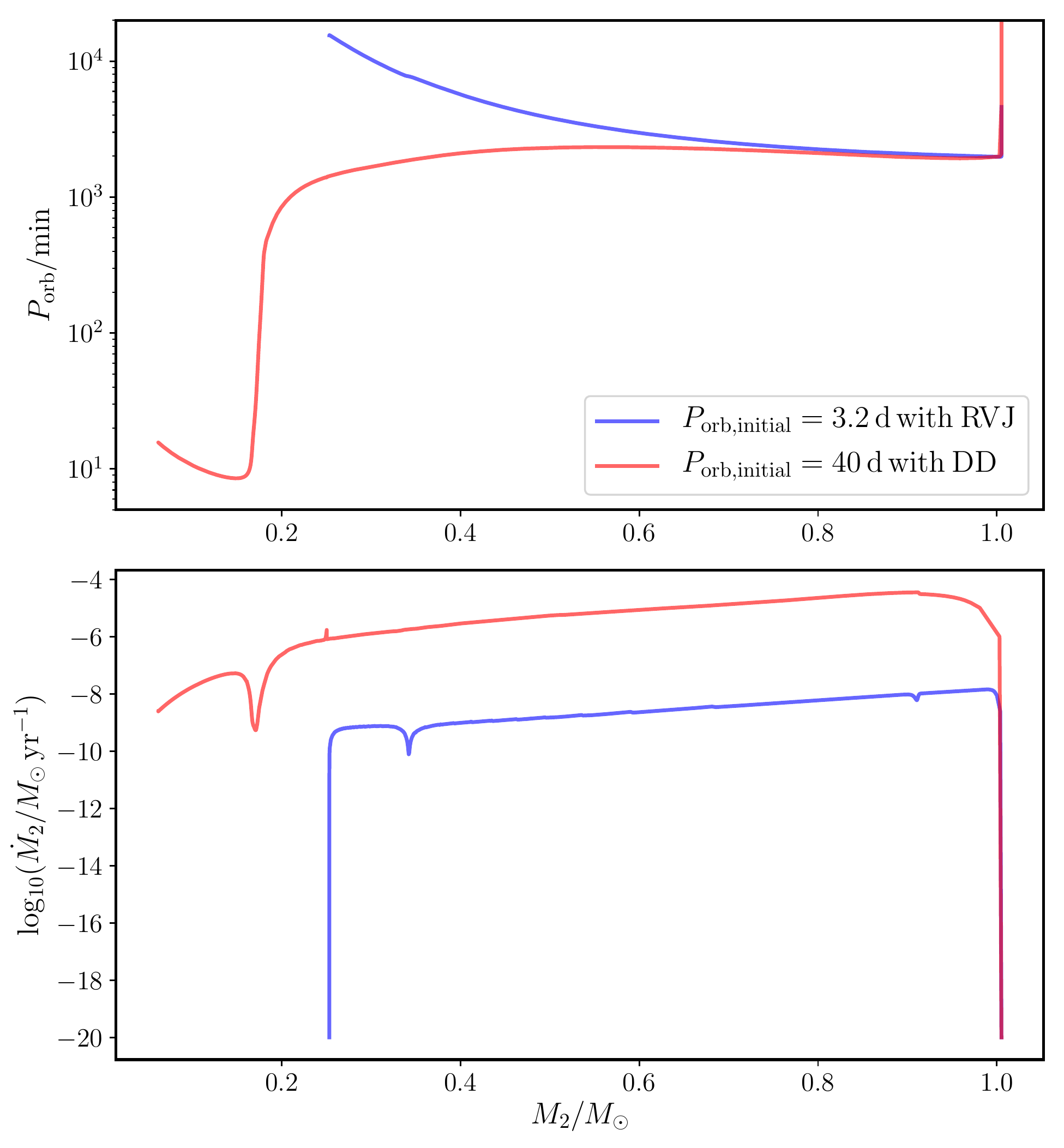}
\caption{The evolution of the donor described in the $(M_2,P_\mathrm{orb})$ and the $(M_2,\mathrm{log_{10}}\dot{M}_2)$ planes for the same systems as in Fig.~\ref{fig:hr-giant}. Under the RVJ model the system ends up as a wide detached binary, whereas under the DD model the system ends up as an AM CVn.}
\label{fig:pm-giant}
\end{figure}

\textcolor{black}{To illustrate the difference in behaviour of the two AML mechanisms when the donor is in its giant phase,} we plot trajectories of systems that commence RLOF after $t_\mathrm{BGB}$ with the DD and the RVJ model in Figs~\ref{fig:hr-giant} and \ref{fig:pm-giant}. In Fig.~\ref{fig:hr-giant}, although RLOF begins at the same point in the HR diagram, the two systems evolve very differently as seen in Fig.~\ref{fig:pm-giant}. The system under the RVJ $\mathrm{AML_{MB}}$ evolves with increasing orbital period and ends up as a wide detached system with $P_\mathrm{orb} \approx 7\,\mathrm{d}$. The initial orbital period or orbital separation which separates converging ultracompact systems to diverging wide systems is known as the bifurcation limit (see \citealt{Nelson2004} and references therein). However the system under the DD $\mathrm{AML_{MB}}$ ends up as an ultracompact CV. This is because, although RLOF commences after the donor ascends the RGB, the high mass-transfer rate ($\Dot{M}_2\approx 10^{-5}M_\odot\,\mathrm{yr}^{-1}$) drives the system faster than its nuclear and thermal time-scale and it ends up as an AM CVn\footnote{For more massive donors the expanding $R_2$ on the RGB and the rapidly shrinking orbit until RLOF lead to dynamical mass transfer and thence common envelope evolution.}. Thus with the DD $\mathrm{AML_{MB}}$ we do not obtain a bifurcation limit. This is a shortcoming of our DD model which can be attributed to the changing degeneracy of the donor's core from the subgiant phase to the red-giant phase. The DD model assumes that the donor's core is radiative and the boundary-layer (BL) dynamo, which drives $\mathrm{AML_{MB}}$, arises because of this boundary \citep[see ST and ][]{Zangrilli1997}. The donor's core is fairly non-degenerate until $t_\mathrm{BGB}$, after which it becomes increasingly degenerate and conductive and the assumption of a radiative core breaks down. This can impede the BL dynamo and lead to a significant decrease in $\mathrm{AML_{MB}}$ in a manner that has not yet been studied. \textcolor{black}{In particular the moment of inertia of the shrinking core falls and it is brought into corotation more easily. This lowers the differential rotation between the core, which was otherwise assumed to be non-rotating, and the envelope, thereby significantly reducing the AML owing to the BL dynamo (equation 15 of ST).} The current implementation of the DD model in the STARS code assumes that any region that is not convective is radiative. This may lead to the observed behaviour. \textcolor{black}{Fixing the overestimation of AML due to the DD model will lead to an overall reduction of the parameter space of AM CVn progenitors. In addition, this change is likely to cause changes in the evolutionary trajectory of the system after the commencement of RLOF, particularly to models where $t_\mathrm{RLOF}\approx t_\mathrm{BGB}$.} Resolving the shortcoming of the DD model around $t_\mathrm{BGB}$ or reproducing the bifurcation limit will be considered in the future.

\subsection{Varying metallicity} 
All the calculations and results here assume a solar metallicity of $Z=0.02$. The underabundance of metals such as iron and silicon has been addressed by low-metallicity progenitors for systems such as GP Com \citep{Marsh1991} and V396 Hya \citep{Kupfer2016}. \cite{Nelson2004} have analysed the metallicity dependence of He-rich donors in binary systems and determined that the bifurcation limit (Section \ref{sec:bif}) is lowered for donors with smaller $Z$. \textcolor{black}{An analysis of the metallicity dependence of our results is beyond the scope of this work and will be considered in the future.}

\section{Conclusion}
\label{sec:Conclusion}
We have investigated in detail the Evolved CV formation mechanism of AM CVn stars using the \textsc{STARS} code. We have shown that the evolution of CVs and their end-products is sensitive to the mechanism of angular momentum loss by the magnetic braking of the donor star. We have used a physically motivated formalism for this magnetic braking mechanism, discussed in detail by \cite{2022MNRAS.513.4169S}, known as the double dynamo (DD) model, and compared it to the empirical magnetic braking formulae of \cite{1983ApJ...275..713R} and its modification by \cite{Knigge2011}. We show that the time-scales of angular momentum loss and subsequent mass loss are shorter with the DD model than with the other two models. Owing to this we find that, with the DD model, a larger parameter space of initial conditions leads to CVs that are the progenitors of AM CVn stars and evolve to form ultracompact CVs within the age of the Galaxy. We make detailed models of these systems and track their evolution beyond their respective period minima and find that evolved CVs populate a region with orbital periods greater than $5.5\,\mathrm{hr}$. We compare our modelled trajectories close to their orbital period minimum and compare the modelled parameters with observations. \textcolor{black}{We find that donors of observed systems such as V396 Hya and YZ LMi can be well modelled with evolved CVs with varying extents of H-exhaustion, and for systems such as CR~Boo and HP~Lib this channel is viable if their actual donor masses lie in the lower-end of the observed mass range.} 

\textcolor{black}{This work does not aim at establishing the Evolved CV channel as a dominant mechanism of AM CVn formation, because even with the increased likelihood of their formation estimated in Section~\ref{sec:probanaly}, this channel is still sub-dominant compared to the other two channels, as shown by \cite{2004MNRAS.349..181N}. In addition, \cite{2023MNRAS.519.2567S} show that most of the AM CVn stars discussed in this work can be explained with the He-star formation channel (see their fig.~7). However, with the work of \cite{Shen2015} suggesting that all WD channel progenitors may merge before forming AM CVn stars, the two remaining viable channels could be just the He-star and the Evolved CV channel. With this work, we have shown that for some of the well know AM CVn stars, the Evolved CV channel is at least a viable possibility. Detailed population synthesis of these binaries with various AML mechanisms can shed more light on better constraining the relative likelihood of AM CVn stars evolving through these two formation channels. We note that similar results could be obtained with other short time-scale magnetic braking mechanisms.}

\section*{Acknowledgements}
\textcolor{black}{The authors thank the anonymous referee for their detailed review of the manuscript. Their helpful comments and suggestions greatly improved the overall structure and content of this work.} AS thanks the Gates Cambridge Trust for his scholarship. HG acknowledges support from NSFC (grant No. 12173081), the key research program of frontier sciences, CAS, No. ZDBS-LY-7005, and Yunnan Fundamental Research Projects (grant No. 202101AV070001). CAT thanks Churchill
College for his fellowship. AS thanks Lev Yungelson for his comments on the manuscript. AS also thanks Alex Hackett for insightful discussions in this field, and Eugene Vasiliev and Daniela Ruz-Mieres for discussions on the kinematics of Gaia14aae.

\section*{Data availability}
The data generated in this work is in Table.~\ref{table:param}.

\bibliographystyle{mnras}
\bibliography{mnras_template} 




 \appendix

\section{Table}
\label{ap:tab}
Table.~\ref{table:param} shows the initial conditions of the systems in which RLOF begins between $t_\mathrm{TMS}$ and $t_\mathrm{BGB}$ for the DD, RVJ and KBP models of $\mathrm{AML_{MB}}$. The region $[a_\mathrm{initial,TMS},\:a_\mathrm{initial,BGB}]$ forms \textcolor{black}{AM CVn stars} with  $a_\mathrm{initial,TMS}$ leading to the least H-exhausted AM CVn donors and $a_\mathrm{initial,BGB}$ leading to the most H-exhausted donors. 

\begin{table*}
\centering
    \begin{tabular}{ |p{2cm}||p{2cm}|p{2cm}||p{2cm}|p{2cm}||p{2cm}|p{2cm}|  }
     \hline
     \multicolumn{7}{|c|}{$M_2=1.0M_\odot$} \\
     \hline
     $M_1/M_\odot$& $a_i(t_\mathrm{TMS,DD})/R_\odot$&$a_i(t_\mathrm{BGB,DD})/R_\odot$&$a_i(t_\mathrm{TMS,KBP})/R_\odot$&$a_i(t_\mathrm{BGB,KBP})/R_\odot$&$a_i(t_\mathrm{TMS,RVJ})/R_\odot$&$a_i(t_\mathrm{BGB,RVJ})/R_\odot$\\
     \hline
     0.6  &  24.28&40.20&   8.71&9.72&  9.38&10.74\\
     0.7  &  23.67&39.59&   8.71&9.55&  9.22&10.74\\
     0.8  &  23.67&38.97&   8.71&9.55&  9.22&10.57\\
     0.9  &  23.06&38.36&   8.71&9.55&  9.22&10.57\\
     1.0  &  23.06&38.36&   8.71&9.55&  9.22&10.57\\
     1.1  &  23.06&38.36&   8.71&9.55&  9.22&10.57\\
     1.2  &  23.06&38.36&   8.71&9.55&  9.22&10.74\\
     \hline   
     
     \hline
     \multicolumn{7}{|c|}{$M_2=1.1M_\odot$} \\
     \hline
     $M_1/M_\odot$& $a_i(t_\mathrm{TMS,DD})/R_\odot$&$a_i(t_\mathrm{BGB,DD})/R_\odot$&$a_i(t_\mathrm{TMS,KBP})/R_\odot$&$a_i(t_\mathrm{BGB,KBP})/R_\odot$&$a_i(t_\mathrm{TMS,RVJ})/R_\odot$&$a_i(t_\mathrm{BGB,RVJ})/R_\odot$\\
     \hline
     0.6  &  23.67&43.26&   8.89&10.11&  9.84&11.61\\
     0.7  &  23.06&42.65&   8.89&10.11&  9.71&11.47\\
     0.8  &  23.06&42.04&   8.76&9.98&  9.71&11.47\\
     0.9  &  22.44&41.42&   8.76&9.98&  9.57&11.47\\
     1.0  &  22.44&41.42&   8.76&9.98&  9.57&11.47\\
     1.1  &  22.44&40.81&   8.76&9.98&  9.57&11.47\\
     1.2  &  22.44&40.81&   8.76&10.11&  9.57&11.47\\
     \hline     
     
     \hline
     \multicolumn{7}{|c|}{$M_2=1.2M_\odot$} \\
     \hline
     $M_1/M_\odot$& $a_i(t_\mathrm{TMS,DD})/R_\odot$&$a_i(t_\mathrm{BGB,DD})/R_\odot$&$a_i(t_\mathrm{TMS,KBP})/R_\odot$&$a_i(t_\mathrm{BGB,KBP})/R_\odot$&$a_i(t_\mathrm{TMS,RVJ})/R_\odot$&$a_i(t_\mathrm{BGB,RVJ})/R_\odot$\\
     \hline
     0.6  &  15.14&46.57&   7.22&10.42&  7.98&12.25\\
     0.7  &  15.14&45.85&   7.06&10.42&  7.83&12.25\\
     0.8  &  14.42&45.14&   7.22&10.42&  7.98&12.10\\
     0.9  &  14.42&45.14&   7.22&10.27&  7.98&12.10\\
     1.0  &  14.42&44.42&   7.22&10.27&  7.98&12.10\\
     1.1  &  14.42&44.42&   7.22&10.27&  7.98&12.10\\
     1.2  &  14.42&43.71&   7.22&10.27&  7.98&12.10\\
     \hline 
     
      \hline
     \multicolumn{7}{|c|}{$M_2=1.3M_\odot$} \\
     \hline
     $M_1/M_\odot$& $a_i(t_\mathrm{TMS,DD})/R_\odot$&$a_i(t_\mathrm{BGB,DD})/R_\odot$&$a_i(t_\mathrm{TMS,KBP})/R_\odot$&$a_i(t_\mathrm{BGB,KBP})/R_\odot$&$a_i(t_\mathrm{TMS,RVJ})/R_\odot$&$a_i(t_\mathrm{BGB,RVJ})/R_\odot$\\
     \hline
     0.6  &  13.42&51.14&   6.45&11.18&  7.22&13.47\\
     0.7  &  12.57&49.42&   6.30&11.03&  7.22&13.22\\
     0.8  &  12.57&49.42&   6.30&11.03&  7.22&13.22\\
     0.9  &  12.57&48.57&   6.30&10.88&  7.06&13.16\\
     1.0  &  12.57&47.71&   6.30&10.88&  7.06&13.16\\
     1.1  &  12.57&47.71&   6.30&10.88&  7.06&13.16\\
     1.2  &  12.57&47.71&   6.30&10.88&  7.06&13.16\\
     \hline 
    \end{tabular}
    \caption{Initial orbital separations for which RLOF begins at the end of the MS $t_\mathrm{TMS}$ and the base of the giant branch $t_\mathrm{BGB}$, denoted by  $a_i(t_\mathrm{TMS})$ and $a_i(t_\mathrm{BGB})$ for various models of  $\mathrm{AML_{MB}}$.}
    \label{table:param}
\end{table*}





\bsp	
\label{lastpage}
\end{document}